\documentclass{article}

\pdfoutput=1

\PassOptionsToPackage{numbers, compress}{natbib}
\usepackage[final]{nips_2017}

\usepackage[utf8]{inputenc} 
\usepackage[T1]{fontenc}    
\usepackage[colorlinks=true,citecolor=blue]{hyperref}       
\usepackage{url}            
\usepackage{booktabs}       
\usepackage{amsfonts}       
\usepackage{nicefrac}       
\usepackage{microtype}      

\usepackage{amsmath}
\usepackage{amsfonts}
\usepackage{mathrsfs}
\usepackage{bm}
\usepackage{dsfont}
\usepackage{stmaryrd}
\usepackage{algorithm}
\usepackage{algorithmic}
\usepackage[T1]{fontenc}    
\usepackage{nicefrac}       
\usepackage{microtype}      
\usepackage{booktabs} 
\usepackage{graphicx}
\graphicspath{{fig/}} 
\usepackage{pdfpages} 
\usepackage{wrapfig}  
\usepackage{footmisc} 

\usepackage{enumitem}
\usepackage{capt-of}
\usepackage{varwidth}

\DeclareMathOperator*{\argmax}{argmax}

\newcommand{\llb}{\llbracket}
\newcommand{\rrb}{\rrbracket}

\newcommand{\eat}[1]{}

\newcommand{\pb}[1]{^{({#1})}}

\newcommand{\bu}{\mathbf{u}}

\newcommand{\x}{\mathbf{x}}
\newcommand{\y}{\mathbf{y}}
\newcommand{\w}{\mathbf{w}}

\newcommand{\eg}{e.g.\ }
\newcommand{\ie}{i.e.\ }

\newcommand{\secmoveup}{\vspace{-3.5mm}}                
\newcommand{\textmoveup}{\vspace{-2.2mm}}               
\newcommand{\itemmoveup}{\vspace{-0.5mm}}                
\newcommand{\eqmoveup}{\vspace{-1.0mm}}                 
\newcommand{\captionmoveup}{\eqmoveup\vspace{-2.80mm}}   

\title{Structured Recommendation}

\author{
    \vspace{5pt} Dawei Chen$^{*\dagger}$ \qquad Lexing Xie$^{*\dagger}$ \qquad Aditya Krishna Menon$^{\dagger*}$ \qquad Cheng Soon Ong$^{\dagger*}$ \\
    \vspace{5pt} $^*$The Australian National University \qquad$^\dagger$Data61, CSIRO \\ 
    \texttt{\{u5708856, lexing.xie\}@anu.edu.au} \\
    \texttt{\{aditya.menon, chengsoon.ong\}@data61.csiro.au} \\
}

\allowdisplaybreaks

\begin{document}

\maketitle

\begin{abstract}

Current recommender systems largely focus on static, unstructured content.
In many scenarios, we would like to recommend content that has structure,
such as 
a trajectory of points-of-interests in a city, or a playlist of songs.
Dubbed {\em Structured Recommendation},
this problem differs from the typical structured prediction problem
in that there are multiple correct answers for a given input.
Motivated by trajectory recommendation, we focus on sequential structures
but in contrast to classical Viterbi decoding we require that
valid predictions are sequences with no repeated elements.
We propose an approach to sequence recommendation based on the structured support vector machine.
For prediction, we modify the inference procedure to avoid predicting loops in the sequence.
For training, we modify the objective function to account for the existence of multiple ground truths for a given input. We also modify the loss-augmented inference procedure to exclude the known ground truths.
Experiments on real-world trajectory recommendation datasets show the benefits of our approach over existing, non-structured recommendation approaches.

\end{abstract}
\captionmoveup


\secmoveup
\section{Introduction}
\label{sec:intro}
\textmoveup

Content recommendation has been the subject of a rich body of literature~\citep{Goldberg:1992,Sarwar:2001,Koren:2010},
with established techniques seeing widespread adoption in industry~\citep{Linden:2003,Agarwal:2013,Amatriain:2015,Gomez-Uribe:2015}.
The success of these methods is explained by both the explosion in availability of user's explicit and implicit preferences for content,
as well as the design of methods that can suitably exploit these to make useful recommendations~\citep{Koren:2009}.

Classical recommendation systems have focused on a fixed set of individual items such as movies, dubbed here as recommending \emph{unstructured} content.
While this setting has considerable value,
in many important scenarios one needs to recommend content that possesses some \emph{structure}.
An example of this is where the desired content is naturally organized in
\emph{sequences} or {\em graphs}.
For example, consider 
recommending a trajectory of points of interest in a city to a visitor~\citep{lu2010photo2trip,lu2012personalized,ijcai15,cikm16paper}, a playlist of songs~\citep{McFee:2011,chen2012playlist,hidasi2015session,choi2016towards},
a chemical compound~\cite{dehaspe1998finding} or a few linked websites for e-commerce~\cite{antikacioglu2015recommendation}.
%

A na\"{i}ve approach to sequence recommendation is to ignore the structure.
In the trajectory example, we could learn a user's preference for individual locations,
and create a trajectory based on the top ranked locations.
However, such an approach may be sub-optimal:
for example,
in the trajectory recommendation problem, it is unlikely a user will want to visit three restaurants in a row.
Similarly,
while a user's two favourite songs might belong
the metal and country genres respectively,
it is questionable that a playlist featuring these songs in succession will be as enjoyable to the user.

The above raises the question of how one can effectively learn from such sequential content.
In this paper, we show how to cast sequence recommendation as a \emph{structured prediction} problem,
which allows us to leverage the toolkit of structured SVMs~\citep{tsochantaridis2005large}.
However, a vanilla application of such methods does not suffice,
as they do not account for the fact that each input can have multiple ground truths,
and that repeated elements in predicted sequences are undesirable.
We show how to overcome this by
incorporating multiple correct sequences into a structured prediction formulation,
and by two novel applications of the list Viterbi algorithm that sequentially finds the list of top-scoring sequences.
Specifically, our contributions are as follows:
\begin{itemize}[noitemsep,leftmargin=12pt]\itemmoveup
	\item We formalise the problem of sequence recommendation (\S\ref{sec:seqrec-defn}), and show how trajectory recommendation can be seen as a special case (\S\ref{sec:trajrec}).

	\item We show how sequence recommendation may be attacked using structured SVMs (\S\ref{sec:recseq}).
	We propose one improvement of structured SVMs to the recommendation problem, so as to account for the existence of multiple ground truths for each input (\S\ref{ssec:sr}).

    \item We propose two novel applications of the list Viterbi algorithm -- an extension of the classic Viterbi algorithm that sequentially finds the list of highest scored sequences under some model --
to exclude multiple ground truths for model learning (\S\ref{ssec:training}),
and to predict sequences without repeated elements, i.e. {\em path}s in state-space (\S\ref{ssec:testing}).

	\item We present experiments on real-world trajectory recommendation problems, and demonstrate our structured recommendation approaches improve over existing non-structured baselines (\S\ref{sec:experiment}).\itemmoveup
\end{itemize}



\secmoveup
\section{The structured recommendation problem}
\label{sec:recseq}
\textmoveup

We introduce the structured recommendation problem that is the focus of this paper.
We then provide some motivating examples, in particular the problem of trajectory recommendation.

\secmoveup
\subsection{Structured and sequence recommendation}
\label{sec:seqrec-defn}

Consider the following general 
\emph{structured recommendation} problem:
given an input query $\x \in \mathcal{X}$ (representing \eg a location, or some ``seed'' song)
we wish to recommend one or more \emph{structured outputs} $\y \in \mathcal{Y}$ (representing \eg a sequence of locations, or songs)
according to a learned \emph{score function} $f(\x,\y)$.
To learn $f$,
we are provided as input a training set
$\{ ( \x\pb{i}, \{ \y\pb{ij} \}_{j=1}^{n_i} ) \}_{i=1}^{n}$,
comprising a collection of inputs $\x\pb{i}$ with an associated \emph{set} of $n_i$ output structures $\{ \y\pb{ij} \}_{j=1}^{n_i}$.

For this work, we assume the output $\y$ is a \emph{sequence} of $l$ points, denoted $y_{1:l}$
where each $y_i$ belongs to some fixed set (\eg places of interest in a city, or a collection of songs).
We call the resulting specialisation the \emph{sequence recommendation} problem,
and this shall be our primary interest in this paper.
In many settings, one further requires the sequences to be \emph{paths}, \ie not contain any repetitions.

As a remark, we note that the assumption that $\y$ is a sequence does not limit the generality of our approach,
as inferring $\y$ of other structures can be achieved using corresponding inference and loss-augmented inference algorithms~\cite{joachims2009predicting}.  

\secmoveup
\subsection{Sequence recommendation versus existing problems}

There are key differences between sequence recommendation and 
standard problems in structured prediction and recommender systems;
this brings unique challenges for both inference and learning.

In a structured prediction problem (for sequences), the goal is to learn from a set of
input vector and output sequence tuples
$\{ (\x\pb{i}, \y\pb{i}) \}_{i = 1}^n$, where
for each input $\x\pb{i}$ there is usually one \emph{unique} output sequence $\y\pb{i}$.
In a sequence recommendation problem, however, we expect that 
for each input $\x\pb{i}$ (\eg users),
there are \emph{multiple} associated outputs
$\{ \y\pb{ij} \}_{j=1}^{n_i}$ (\eg trajectories visited).
Structured prediction approaches do not have a standard way to handle such multiple output sequences.
Furthermore, it is desirable for the recommended sequence to consist of unique elements,
or be a {\em path}~\cite{west2001introduction} in the candidate space (\eg locations).
Classic structured prediction does not constrain the output sequence, and having such a
path constraint makes both inference and learning harder.

In a typical recommender systems problem, the outputs are non-structured; canonically, one works with {static} content such as books or movies~\citep{Goldberg:1992,Sarwar:2001,Netflix}.
Thus, making a prediction involves enumerating all {\em non-structured} items $y$ in order to compute $\argmax_y f(\x,y)$ for suitable score function $f$, \eg some form of matrix factorisation~\citep{Koren:2009}.
For sequence recommendation, computing $\argmax_\y f(\x,\y)$ is harder since it is often impossible to efficiently enumerate $\y$ (\eg all possible trajectories in a city).
This inability to enumerate $\y$ also poses a challenge in designing a suitable $f(\x,\y)$,
\eg
matrix factorisation
would require associating a latent feature with each $\y$, which will be infeasible.

\secmoveup
\subsection{Examples of sequence recommendation}
\label{sec:trajrec}

To make the sequence recommendation problem more concrete,
we provide three specific examples,
starting with the problem of trajectory recommendation
that shall serve as a recurring motivation.
Note that in all these problems, one is specifically interested in sequences that are paths.




\textbf{Trajectory recommendation}.
Travel recommendation problems involve a set of points-of-interest (POIs) $\mathcal{P}$ in a city~\cite{bao2015recommendations,zheng2015trajectory,zheng2014urban}.
Given a \emph{trajectory query} $\mathbf{x} = (s, l)$,
comprising a start POI $s \in \mathcal{P}$ and trip length
$l \!>\! 1$ (\ie the desired number of POIs, including $s$),
the \emph{trajectory recommendation} problem is to
recommend one or more sequences of POIs 
that maximise some notion of utility,
learned from a training set
of trajectories visited by travellers in the city.

%


%

Existing approaches treat the problem as one of determining a score for the intrinsic appeal of each POI.
For example, a RankSVM model
which 
learns to predict whether a POI is likely to appear ahead of another POI in a trajectory corresponding to some query~\cite{cikm16paper}.
Formally,
from the given set of trajectories
we derive a training set
$\{ ( \x^{(i)}, \mathbf{r}^{(i)} ) \}_{i = 1}^n$,
where for each trajectory query $\x^{(i)}$ there is a list of ranked POI pairs
$\mathbf{r}^{(i)} \subseteq \mathcal{P} \times \mathcal{P}$
such that
$(p, p') \in \mathbf{r}^{(i)}$
iff
POI $p$ appears more times than POI $p'$ in all trajectories associated with $\x^{(i)}$. 
The training objective is then
\begin{align}
\resizebox{0.7\linewidth}{!}{$\displaystyle
\displaystyle \min_{\w} ~\frac{1}{2} \w^\top \w + C \cdot \sum_{i = 1}^n \sum_{(p, p') \in \mathbf{r}^{(i)}}
\ell\left( \w^\top ( \Psi( \x^{(i)}, p ) - \Psi( \x^{(i)}, p' ) ) \right),
$} \label{eq:ranksvm}
\end{align}
for
feature mapping $\Psi$,
regularisation strength $C$ 
and squared hinge loss $\ell( v ) = \max( 0, 1 - v )^2$.

We can view trajectory recommendation as sequence recommendation in the following way:
given trajectory query $\x$, and a suitable scoring function $f$, we wish to find
$\mathbf{y}^* = \argmax_{\mathbf{y} \in \mathcal{Y}}~f(\mathbf{x}, \mathbf{y}),$
where $\mathcal{Y}$ is the set of all possible trajectories with POIs in $\mathcal{P}$ that conform to the constraints imposed by the query $\mathbf{x}$.
In particular,
$\mathbf{y} = (s,~ y_2, \dots, y_l)$ is a trajectory with $l$ POIs. 
This was the view proposed in~\cite{cikm16paper} where they authors considered an
objective function that added two components together: a POI score and a transition score.

Now, our training set of historical trajectories may be written as
$\{ ( \x\pb{i}, \{ \y\pb{ij} \}_{j=1}^{n_i} ) \}_{i=1}^{n}$,
where each $\x\pb{i}$ is a distinct query
with $\{ \y\pb{ij} \}_{j=1}^{n_i}$ the corresponding \emph{set} of observed trajectories.
Note that we expect most queries to have several distinct trajectories;
minimally,
for example,
there may be two nearby POIs that are visited in interchangeable order by different travellers.
We are also interested in predicting paths $\y$, since it is unlikely a user will want to visit the same location twice.

\textbf{Playlist generation}.
As another example, consider recommending song playlists (\ie sequences) to users, given a query song~\citep{McFee:2011,chen2012playlist,hidasi2015session,choi2016towards}.
A canonical approach is to
learn a latent representation of songs from historical playlists,
and exploit a Markovian assumption on the song transitions.

%
\textbf{Next basket and next song prediction}.
One more example of sequence recommendation is the problem of recommending the next items a user might like to purchase, given the sequence of their shopping basket purchases~\citep{Rendle:2010,Wang:2015}.
The canonical approach here is to apply matrix factorisation to the Markov chain of transitions between items.
This method is feasible because one is only interested in predicting sequences one element at a time, instead of predicting the entire sequence given the initial item.
Recent approaches using recurrent neural networks for
next basket prediction~\cite{yu2016dynamic} and playlist
generation~\cite{choi2016towards} also focus on modelling high quality transitions only.

\subsection{The case for exploiting structure}

Each of the sequence recommendation problems above can be plausibly solved with approaches that do not exploit the structure inherent in the outputs $\y$. 
While such approaches can certainly be useful,
their modelling power is inherently limited,
as
they cannot ensure the \emph{global} cohesion of the corresponding recommendations $\y$.
For example, in the trajectory recommendation problem, the RankSVM model 
might find three restaurants to be the highest scoring POIs;
however, it is unlikely that these form a trajectory that most travellers will enjoy.

This motivates an approach to sequence recommendation that directly ensures such global cohesion.
We now see how to do so with novel extensions to structured prediction approaches. 



%

%




\secmoveup
\section{A structured prediction approach to sequence recommendation}
\label{sec:recseq}
\textmoveup

We now describe a method to
solve sequence recommendation problems.
We first provide background on structured SVMs (SSVM) (Sec~\ref{ssec:ssvm}),
then present a model for structured recommendation (Sec~\ref{ssec:sr}),
followed by the correspondingly updated algorithms for its learning (Sec~\ref{ssec:training})
and inference (Sec~\ref{ssec:testing}).

\secmoveup
\subsection{Background: SSVM for structured prediction}
\label{ssec:ssvm}
\textmoveup

A structured prediction task involves predicting some structured label $\y \in \mathcal{Y}$ for an input $\x \in \mathcal{X}$,
typically via a score function $f(\x,\y)$ that determines the affinity of an (input, label) pair.
A popular model is the Structured Support Vector Machine (SSVM)~\cite{joachims2009predicting,tsochantaridis2005large}, comprising three core ingredients.

\emph{Score function}. In SSVMs, we specify that the score function $f(\x, \y)$ takes a linear form, \ie
$f(\x, \y) = \w^\top \Psi(\x, \y)$,
where $\w$ is a weight vector, and $\Psi(\x, \y)$ is a \emph{joint feature map}.

The design of the feature map $\Psi$ is problem specific.
For sequence prediction problems,
consider the unary
terms for each element in the label $y_{1:l}$, and pairwise interactions between
adjacent elements in the label,
i.e. $y_j$ and $y_{j+1}$ for $j=1 : l \!-\! 1$.
Subsequently, $f(\x,\y)$ decomposes into a weighted sum of
each of these features: 
\begin{equation}
\label{eq:jointfeature}
f(\x, \y) = \w^\top \Psi(\x, \y) =
\sum_{j=1}^l \w_j^\top \psi_j(\x, y_j) + \sum_{j=1}^{l-1} \w_{j,j+1}^\top \psi_{j,j+1}(\x, y_{j}, y_{j+1}).
\eqmoveup
\end{equation}
Here, $\psi_j$ is a feature map between the input and one output label element $y_j$, with a corresponding weight vector $\w_j$,
and $\psi_{j,j+1}$ is a pairwise feature vector that captures the interactions between consecutive labels $y_j$ and $y_{j+1}$,
with a corresponding weight vector $\w_{j,j+1}$.

\emph{Objective function}.
To learn a suitable weight vector $\w$, SSVMs solve: 
\begin{equation}
\label{eq:nslack}
\resizebox{0.93\linewidth}{!}{$\displaystyle
\begin{aligned}
\min_{\w, \, \bm{\xi} \ge 0} ~\frac{1}{2} \w^\top \w + \frac{C}{n} \sum_{i=1}^n \xi_i,  ~~s.t.~  \forall i,
  \w^\top \Psi(\x^{(i)}, \y^{(i)}) - \w^\top \Psi(\x^{(i)}, \bar{\y}) \ge
  \Delta(\y^{(i)}, \bar{\y}) - \xi_i, \, \forall \bar\y \in \mathcal{Y}.
\end{aligned}
$} \eqmoveup
\end{equation}

Here,
$\mathcal{Y}$ is the set of all possible sequences
and $\Delta(\y\pb{i}, \bar\y)$ is the loss function between $\bar\y$ and the ground truth $\y\pb{i}$,
and slack variable $\xi_i$ is the \emph{hinge loss} for the prediction of the $i$-th example~\cite{tsochantaridis2005large}.
Alternatively, we can use \emph{one} slack variable to represent the sum of the $n$ hinge losses in~(\ref{eq:nslack}),
\ie the 1-slack formulation~\cite{tsochantaridis2005large}.

\emph{Loss-augmented inference}.
We can rewrite the constraints in (\ref{eq:nslack}) as
$\w^\top \Psi(\x\pb{i}, \y\pb{i}) + \xi_i \ge
\max_{\bar\y \in \mathcal{Y}} \left\{ \Delta(\y\pb{i}, \bar\y) + \w^\top \Psi(\x\pb{i}, \bar\y) \right\}$,
and the maximisation at the right side is known as the loss-augmented inference.
When the underlying graph of SSVM is a tree (which is the case for sequence recommendation),
this may be done efficiently if the loss function $\Delta(\cdot,\cdot)$ is decomposable
with respect to individual and pairs of label elements,
\eg using the Viterbi algorithm~\cite{joachims2009predicting}.

\secmoveup
\subsection{SSVM for structured recommendation: the SP and SR models}
\label{ssec:sr}
\textmoveup

We now present two possible means of applying SSVMs to sequence recommendation.

\emph{The SP model}.
Recall that structured recommendation
involves observing \emph{multiple} ground truth outputs for each input, \ie
for input $\x\pb{i}$, there is a set of ground truths $\{\y\pb{ij}\}_{j=1}^{n_i}$.
One na\"{i}ve approach to the problem
is creating
$n_i$ examples $\{(\x\pb{i}, \y\pb{ij})\}_{j=1}^{n_i}$,
and feeding this to the classic SSVM. 
We call this the \emph{structured prediction} (\emph{SP}) model.

The SP model is appealing due to its simplicity.
However, it is sub-optimal:
the result of loss-augmented inference on $(\x\pb{i}, \y\pb{ij})$ could be a ground truth label $\y' \in \{\y\pb{ij}\}_{j=1}^{n_i}$,
which means we would incorrectly penalise predicting $\y'$ for $\x\pb{i}$.

\emph{The SR model}.
To overcome the limitation of the SP model,
we propose the following \emph{structured recommendation} (\emph{SR}) model that extends the SSVM to explicitly incorporate multiple ground truths: 
\begin{equation}
\label{eq:nslack_ml}
\resizebox{0.93\linewidth}{!}{$
\begin{aligned}
\min_{\w, \, \bm{\xi} \ge 0} ~ \frac{1}{2} \w^\top \w + \frac{C}{N} \sum_{i=1}^n \sum_{j=1}^{n_i} \xi_{ij}, ~~s.t.~ \forall i, \, \forall j,
  \w^\top \Psi(\x^{(i)}, \y^{(ij)}) - \w^\top \Psi(\x^{(i)}, \bar{\y}^{(i)}) \ge
  \Delta(\y^{(ij)}, \bar{\y}^{(i)}) - \xi_{ij}.
\end{aligned}
$}
\end{equation}
where $N = \sum_i n_i$ and $\bar{\y}^{(i)} \in \mathcal{Y} \setminus \{\y^{(ij)}\}_{j=1}^{n_i}$.
The 1-slack formulation of~(\ref{eq:nslack_ml}) can be formed similar to that of~(\ref{eq:nslack}).
Intuitively, the objective in (\ref{eq:nslack_ml}) is similar to a ranking objective, as the constraints enforce
that the positively labeled sequences (the known items that the user likes) are scored
higher than all other unseen sequences.
Such objectives have been proven useful in classic unstructured recommendation~\cite{bpr09}.

Compared to the SP model (\ref{eq:nslack}), the key distinction is that (\ref{eq:nslack_ml})
explicitly aggregates all the ground truth sequences for each input when generating the constraints,
\ie the loss-augmented inference becomes
\begin{equation}
\label{eq:loss_aug_inf}
\max_{\bar{\y}^{(i)} \in \ \mathcal{Y} \setminus \{\y^{(ij)}\}_{j=1}^{n_i}}
     \left( \Delta(\y^{(ij)}, \bar{\y}^{(i)}) + \w^\top \Psi(\x^{(i)}, \bar\y^{(i)}) \right).
\end{equation}
In this way, we do not have contradictory constraints where
two ground truth sequences are each required to have larger score than the other.


The SP and SR models can be learned using a rich set of training schemes such as
cutting-plane algorithms~\cite{joachims2009predicting}, 
gradient-based algorithms~\cite{ratliff2006subgradient} 
and conditional gradient methods~\cite{lacoste2013block} 
giving proper loss augmented inference and prediction procedures,
which we describe in the next two sections.

Beyond dealing with multiple ground truths,
one additional requirement is for the predicted sequence to be a {\em path}.
This is a {\em global} constraint bringing new challenges for prediction and loss-augmented inference.
The following two sections describe the training and prediction with these requirements.

\secmoveup
\subsection{SP and SR model training}
\label{ssec:training}
\textmoveup

The main challenge in training the SP and SR models is performing loss-augmented inference~(\ref{eq:loss_aug_inf}).
As noted above, the SP model can be trained as per the vanilla SSVM.
The SR model, however, requires modifying the training procedure to account for the existence of multiple ground truths.
In particular, (\ref{eq:loss_aug_inf}) needs to be solved by \emph{excluding the sequences} $\{\y\pb{ij}\}_{j=1}^{n_i}$, and ideally with {\sc path} constraints.
We show in the following how to address both problems with an extension of the Viterbi algorithm.


\textbf{The list Viterbi algorithms}.
List Viterbi represents a family of algorithms originally invented to decode digital signals corrupted by noise~\cite{seshadri1994list,nill1995list}, or to find more than one candidate sentence in speech recognition~\cite{soong1991tree}. Given a score function that can be decomposed into unary and pairwise costs such as (\ref{eq:jointfeature}), a list Viterbi algorithm aims to find the $k$ highest scoring sequences. 

The \emph{parallel list Viterbi} algorithm~\cite{seshadri1994list} finds the top $k$ sequences
by keeping $k$ backtrack pointers for each possible state in each position of the sequence.
This algorithm is memory-inefficient and can be difficult to use when one does not know what $k$ to use apriori -- this is the case for both the learning and inference challenges in this work.

An important variant is the \emph{serial list Viterbi} algorithms (SLVA)~\cite{seshadri1994list,nill1995list,nilsson2001sequentially}.
This algorithm sequentially find the $k$-th best sequences given the best, 2nd best, \dots, $(k \!-\! 1)$-th best sequences.
The key insight here is that the 2nd best sequence deviates from the best sequence
for one continuous segment, and then finally merges back to the best sequence without deviating again
-- otherwise replacing one of the continuous segments with those from the best sequence will increase the score.
Subsequently, the $k$th best sequence can be the 2nd best sequence relative to the $(k \!-\! 1)$-th sequence
at the point of final merge, or the 2nd or 3rd best sequence to the $(k \!-\! 2)$-th sequence at the point of final merge, \ldots, and so on. SLVA works by keeping track of the score differences to the best sequences and merge points along the sequences. See supplement for a complete description.

We use SLVA in two different ways, first to deal with multiple ground truths, and second
to find paths.

\textbf{Eliminating multiple ground truths}.
Recall that standard loss-augmented inference for an SSVM (for sequences) may be done with the classic Viterbi algorithm, but the best-scoring sequence could be in the ground-truth set $\{\y^{(ij)}\}_{j=1}^{n_i}$. We check whether this is the case, and if it is, use SLVA to decode the next best sequence until we find one that is not in the ground truth set.
Note that the serial list Viterbi algorithm
can be used for loss-augmented inference with Hamming loss, the most common loss function for sequence prediction tasks,
since it only requires the loss function be decomposable with respect to the label elements.

\textbf{Eliminating loops: the \textsc{SPpath} and \textsc{SRpath} model}.
The list Viterbi algorithm is also applicable to enforce that loss-augmented inference only considers sequences that are \emph{paths}. This may be done by checking if each of the next-best sequences has a loop, and following SLVA as above.
This idea can be applied to both the SP and SR models, as enforcing path constraints is independent of excluding multiple ground truths.
We call these extended models \textsc{SPpath} and \textsc{SRpath} respectively.

Note that $k$ is expected to be small in noisy signal recovery, such as digital communications~\cite{nill1995list} and speech recognition~\cite{soong1991tree}. But this is not necessarily the case for inference in SR models.

\secmoveup
\subsection{SP and SR model prediction}
\label{ssec:testing}
\textmoveup

For both the SP and SR models, prediction requires that we compute $\argmax_{\y \in \mathcal{Y}_{\mathrm{path}}} f( \x, \y )$.
This is the maximum over $\mathcal{Y}_{\mathrm{path}} \subseteq \mathcal{Y}$,
which comprises all elements of $\mathcal{Y}$ that are paths and not all possible sequences.
Observing that this requires excluding a set of sequences from consideration, a natural idea is to apply the \emph{list Viterbi algorithm} of the previous section.
We further observe that our {\em path-decoding} problem is a variant of the well-known
{\em traveling salesman problem} (TSP), in particular the best of ${m \choose l}$ TSPs.
This is because list Viterbi can find all best-scoring sequences until one of which is a path, while TSP can be formulated to find the best path.
This equivalence relation opens new possibilities to leverage well-studied TSP formulations
and modern implementations~\cite{tspbook2011}. In particular, we consider an integer linear program (ILP) formulation of the TSP (Eq.~\ref{eq:obj} to \ref{eq:cons4}). This formulation extends earlier work on sequence recommendation~\cite{lian2014geomf,cikm16paper} by systematically incorporating unary and pairwise scoring terms.



Given a set of points $\{p_j\}_{j=1}^m$,
consider binary decision variables $u_{jk}$ that are true if and only if
the transition from $p_j$ to $p_k$ is in the resulting path,
and binary decision variables $z_j$ that are true iff
the optimal path $\y^*$ terminates at $p_j$.
For brevity, we index the points such that $y_1^* = p_1$.
Firstly, the desired path should start from $y_1^*$ (Constraint~\ref{eq:cons1}).
In addition, only $l\!-\!1$ transitions between points are permitted as the path length is $l$ (Constraint~\ref{eq:cons2}).
Moreover, any point could be visited at most once (Constraint~\ref{eq:cons3}).
The last constraint, where $v_j \in \mathbf{Z}$ is an auxiliary variable,
enforces that $\y^*$ is a single sequence of points without sub-tours.
We rewrite Eq.~(\ref{eq:jointfeature}) into a linear function of decision variables $u_{jk}$
(dropping the unary term corresponding to $y_1^*$ as it is a constant), which results in ~(\ref{eq:obj}).
\setlength{\belowdisplayskip}{1pt} \setlength{\belowdisplayshortskip}{0pt}
\setlength{\abovedisplayskip}{1pt} \setlength{\abovedisplayshortskip}{0pt}
\begin{equation}
\label{eq:obj}
\max_{\bu} ~\sum_{k=1}^m \w_k^\top \psi_k(\x, p_k) \sum_{j=1}^m u_{jk} +
            \sum_{j,k=1}^m u_{jk} \w_{jk}^\top \psi_{j, k}(\x, p_j, p_k)
\end{equation}
{$\displaystyle
\begin{minipage}{0.45\linewidth}
\begin{alignat}{2}
s.t. \,
& \sum_{k=2}^m u_{1k} = 1, \, \sum_{j=2}^m u_{j1} = z_1=0                 \label{eq:cons1} \\
& \sum_{j=1}^m \sum_{k=1}^m u_{jk} = l-1, \, \sum_{j=1}^m u_{jj}=0        \label{eq:cons2}
\end{alignat}
\end{minipage}
\qquad
\begin{minipage}{0.5\linewidth}
\begin{alignat}{3}
\sum_{j=1}^m u_{ji} = z_i &+ \sum_{k=2}^m u_{ik} \le 1, \, i = 2,\cdots,m  \label{eq:cons3} \\
v_j - v_k + 1 &\le (m-1) (1-u_{jk}),                                       \nonumber \\
              & \qquad j,k = 2,\cdots,m                                    \label{eq:cons4}
\end{alignat}
\end{minipage}
$}

\textbf{Recommending more than one trajectory}.
Since multiple possible trajectories can start at the same POI, 
we need to predict multiple trajectories for each test query. 
We can use the above ILP formulation to find the top-$K$ scored paths in a sequential manner.
In particular, given the $K\!-\!1$ top scored paths $\{\y\pb{k}\}_{k=1}^{K-1}$,
the $K\!$-th best scored path can be found by solving the above ILP with additional constraints:
$\sum_{j=1}^{l-1} u_{y_j, y_{j+1}} \le l-2, ~\forall \y \in \{\y^{(k)}\}_{k=1}^{K-1}$.
Alternatively, we can achieve this by using the same approach as when eliminating ground truth sequences using SLVA.

\textbf{Eliminating multiple ground truths with ILP?}
A natural question is whether one can use the above ILP
to exclude observed labels when training an SR model,
\ie solving the loss-augmented inference~(\ref{eq:loss_aug_inf}).
In fact, this may be done
as long as the loss $\Delta(\y,\bar\y)$ can be represented as a linear function of $u_{jk}$,
\eg the number of mis-predicted POIs disregarding the order $\Delta(\y, \bar\y) = \sum_{j=2}^l (1 - \sum_{k=1}^m u_{k, y_j})$.
However, we note that Hamming loss 
cannot be used here, as $\Delta(\y,\bar\y) = \sum_{j=1}^l (1 - \llb y_j = \bar y_j \rrb)$
cannot be expressed as a linear function of $u_{jk}$.

\textbf{Practical choices: ILP vs SLVA vs other heuristics}.
When performing prediction for SP and SR model, we found that state-of-the-art ILP solvers converge to a solution faster
than the serial list Viterbi algorithm if the sequence length $l$ is large.
The reason is, while list Viterbi algorithms are polynomial time given the list depth $k$~\cite{nilsson2001sequentially},
in reality $k$ is unknown a priori and can be very large for long sequences.
We therefore use ILP for very long ($l\ge10$) sequences in the experiments (otherwise the list Viterbi algorithm is applied).

One might also consider the well-known Christofides algorithm~\cite{christofides1976} to eliminate loops in a sequence, as this is known to generate a solutions within a factor of 3/2 of the optimal solution for traveling salesman problems. However, the resulting sequence will have less than the desired number of points, and the resulting score will not be optimal.

\secmoveup
\section{Experimental results}
\label{sec:experiment}
\textmoveup

We present empirical evaluations for the trajectory recommendation task in
Section~\ref{sec:trajrec}.
Results are reported on real-world datasets of photo tours
created from the publicly available YFCC100M corpus~\cite{thomee2016yfcc100m}.

\secmoveup
\subsection{Photo trajectory datasets}
\label{sec:dataset}
\textmoveup

We used the trajectory data\footnote{\url{https://bitbucket.org/d-chen/tour-cikm16}}
extracted from Flickr photos for the cities of Glasgow, Osaka and
Toronto~\cite{ijcai15,cikm16paper}.
Each dataset comprises of a
list of trajectories, being a sequence of points of interest (POI),
as visited by various Flickr users and recorded by the geotags in photos.
Table~\ref{tab:data} summarises the profile of each dataset.
We see that most queries have more than one ground truth, making the sequence recommendation setting relevant. Further, each query has an average of 4-9, and a maximum of 30-60 trajectories (details in supplement).
In all datasets,
each user has on average less than two trajectories.
This makes user-specific recommendation impractical, and also undesirable because
a user would want different recommendations given different starting locations, and not a static recommendation no matter where she is.
The sparsity of this dataset presents a barrier for large-scale evaluations.
Music playlist datasets are larger, but recent results show that sequencing information does not affect the data likelihood~\cite{chen2012playlist}.

\begin{table}[t]
	\begin{minipage}[t]{\linewidth}
		\resizebox{\linewidth}{!}{
		\setlength{\tabcolsep}{4pt} 
		\small
		\begin{tabular}{lllll|ccc|cc} \hline 
		\textbf{Dataset} & \textbf{\#Traj} & \textbf{\#POIs} & \textbf{\#Users} & \textbf{\#Queries} & \textbf{\#GT=1} & \textbf{\#GT$\in [2,5]$} & \textbf{\#GT$>$5} & \textbf{\#shortTraj} & \textbf{\#longTraj} \\ \hline
		Glasgow          & 351              & 25              & 219              & 64                 & 23              & 22                      & 19                & 336                     & 15 \\
		Osaka            & 186              & 26              & 130              & 47                 & 17              & 22                      & 8                 & 178                     & 8  \\
        Toronto          & 977              & 27              & 454              & 99                 & 30              & 33                      & 36                & 918                     & 59 \\
		\hline
		\end{tabular}%
		}
		\captionof{table}{Statistics of trajectory datasets.
        Including the number of trajectories (\#Traj), POIs (\#POIs), users (\#Users), queries (\#Queries);
        the number of queries with a single (\#GT=1), 2-5 (\#GT$\in$[2,5]), or more than 5 (\#GT$>$5) ground truths;
        and profile of trajectory length, \ie less than 5 (\#shortTraj) and more than 5 POIs (\#longTraj).
        }
		\label{tab:data}
	\end{minipage}\captionmoveup
\end{table}




\secmoveup
\subsection{Evaluation settings}
\label{ssec:methods}
\textmoveup

We compare the performance of our methods to the following three baselines:
\begin{itemize}[leftmargin=0.125in]\itemmoveup
\parskip -.05em
\item The \textsc{Random} baseline recommends a sequence of POIs by sampling uniformly at random from the whole set of POIs
      (without utilising any POI or query related features). To obtain top-$k$ predictions
			we independently repeat $k$ times.

\item The stronger \textsc{Popularity} baseline recommends the top-$l$ most popular POIs,
      \ie, the POIs visited by the most number of users in the training set.

\item \textsc{PoiRank}~\cite{cikm16paper}
      considers a number of POI-query features (see supplement) in addition to the popularity,
      and trains a RankSVM model~\eqref{eq:ranksvm} to learn a score for each POI. The top-$l$ scored POIs are then used to construct a trajectory.
\end{itemize}\itemmoveup
To perform top-$k$ prediction with \textsc{Popularity} and \textsc{PoiRank},
we make use of the same approach we used to deal with multiple ground truths.
For \textsc{Popularity}, the score of a path is the accumulated popularity of all POIs in the path;
for \textsc{PoiRank}, the score of a path is the likelihood
(the ranking scores for POIs are first transformed into a probability distribution using the softmax function, as described in~\cite{cikm16paper}).
We consider four variants of sequence recommendation, starting with a structured prediction model, then incorporating multiple ground truths, and finally path constraints:
\begin{itemize}[leftmargin=0.125in]
\item The SP and SR methods, described in Section~\ref{ssec:sr}, using both POI-query features and pairwise features (see supplement).

\item {\sc SPpath} and {\sc SRpath}, described in Section~\ref{ssec:training} with the same features as SP and SR.
\end{itemize}\itemmoveup


We evaluate each algorithm using leave-one-query-out cross validation.
That is, holding out all the relevant trajectories for each query $\x\pb{i}$ (\ie $\{\y\pb{ij}\}_{j=1}^{n_i}$) in each round.
The regularisation constant $C$ is tuned using Monte Carlo cross validation~\cite{burman1989comparative} on the training set.
We use three performance measures for POIs, sequences and ordered lists.
The {\bf F$_1$ score on points}~\cite{ijcai15} computes F$_1$ on the predicted versus seen points
without considering their relative order.
The {\bf F$_1$ score on pairs}~\cite{cikm16paper} is proposed to mitigate this by computing F$_1$ on all ordered pairs in the predicted versus ground truth sequence. 
The well-known rank correlation {\bf Kendall's $\tau$}~\cite{agresti2010analysis}
computes the ratio of concordant (correctly ranked) pairs minus discordant pairs, over all possible pairs after accounting for ties.

Structured recommendation performs ranking on a very large labelset (of size $m^l$).
We report results on the {\em best of top $k$}~\cite{russakovsky2015imagenet}.
That is, for all methods described in Section~\ref{ssec:methods},
We predict the top $k$ trajectories\footnote{To get $k$ paths, the list Viterbi algorithm normally searches a long list which contains sequences with loops.}
and then report the best match of any in the top $k$ to any trajectory in the ground truth set.
We reiterate that irrespective of the training procedure, \textsc{SP}, \textsc{SR},
\textsc{SPpath} and \textsc{SRpath} all use prediction procedures that eliminate subtours.

\eat{
As described previously, our methods are capable of recommending not merely a single trajectory,
but rather a list of trajectories.
While one can take the top recommended trajectory as the prediction,
this ignores the fact that there are likely multiple plausible trajectories for any given query.
Thus, for each performance measure $\mathrm{perf}$,
we take the maximum over all trajectories,
i.e.,
\begin{equation*}
\mathrm{perf}^{(i)}( \mathbf{y}, \hat{\mathbf{y}} ) =
\max_{(\mathbf{y}, \hat{\mathbf{y}}) \in \{\mathbf{y}^{(ij)}\}_{j=1}^{N_i} \times \{\hat{\mathbf{y}}^{(ij)}\}_{j=1}^k}
\mathrm{perf}(\mathbf{y}, {\hat{\mathbf{y}}}),
\end{equation*}
where $\{\mathbf{y}^{(ij)}\}_{j=1}^{N_i}$ are the ground truths for query $\mathbf{x}^{(i)}$ and
$\{\hat{\mathbf{y}}^{(ij)}\}_{j=1}^k$ are the top-$k$ recommendations.
}

\secmoveup
\subsection{Results and discussion}
\label{sec:result}
\textmoveup


\begin{table*}[t]\captionmoveup
     \caption{Results on trajectory recommendation datasets on best of top-10.
     Higher scores are better for all metrics. Bold entries: \textbf{best} performing method for each metric; italicised entries: the \textit{second best}.
     }
     \label{tab:result}
     \centering
     \resizebox{\linewidth}{!}{
\begin{tabular}{l|cc|cc|ccc} \hline
& \multicolumn{7}{c}{\bf Kendall's $\tau$} \\ \hline
 & \textsc{Random} & \textsc{Popularity} & \textsc{PoiRank} & \textsc{SP} & \textsc{SPpath} & \textsc{SR} & \textsc{SRpath} \\ \hline
Glasgow & $0.703\pm0.029$ & $0.748\pm0.036$ & $0.830\pm0.029$ & $0.790\pm0.030$ & $0.787\pm0.029$ & $\mathbf{0.868\pm0.026}$ & $\mathit{0.853\pm0.026}$ \\
Osaka & $0.685\pm0.035$ & $0.768\pm0.038$ & $0.787\pm0.037$ & $0.749\pm0.043$ & $\mathit{0.791\pm0.036}$ & $0.777\pm0.036$ & $\mathbf{0.803\pm0.034}$ \\
Toronto & $0.652\pm0.024$ & $0.719\pm0.024$ & $0.784\pm0.023$ & $0.697\pm0.027$ & $0.719\pm0.026$ & $\mathbf{0.802\pm0.022}$ & $\mathit{0.797\pm0.022}$ \\
\hline
& \multicolumn{7}{c}{\bf F$_1$ score on points} \\ \hline
Glasgow & $0.731\pm0.026$ & $0.771\pm0.033$ & $0.847\pm0.025$ & $0.810\pm0.027$ & $0.807\pm0.026$ & $\mathbf{0.883\pm0.023}$ & $\mathit{0.868\pm0.023}$ \\
Osaka & $0.703\pm0.032$ & $0.786\pm0.034$ & $0.804\pm0.034$ & $0.770\pm0.039$ & $\mathit{0.809\pm0.033}$ & $0.793\pm0.033$ & $\mathbf{0.820\pm0.031}$ \\
Toronto & $0.696\pm0.021$ & $0.746\pm0.022$ & $0.807\pm0.020$ & $0.733\pm0.023$ & $0.755\pm0.022$ & $\mathbf{0.828\pm0.019}$ & $\mathit{0.823\pm0.020}$ \\
\hline
& \multicolumn{7}{c}{\bf F$_1$ score on pairs} \\ \hline
Glasgow & $0.495\pm0.046$ & $0.623\pm0.051$ & $0.726\pm0.043$ & $0.658\pm0.046$ & $0.648\pm0.045$ & $\mathbf{0.770\pm0.039}$ & $\mathit{0.746\pm0.041}$ \\
Osaka & $0.451\pm0.057$ & $0.626\pm0.055$ & $0.661\pm0.056$ & $0.620\pm0.061$ & $\mathit{0.664\pm0.055}$ & $0.637\pm0.055$ & $\mathbf{0.671\pm0.053}$ \\
Toronto & $0.438\pm0.034$ & $0.550\pm0.035$ & $0.649\pm0.033$ & $0.530\pm0.037$ & $0.552\pm0.036$ & $\mathbf{0.660\pm0.033}$ & $\mathit{0.657\pm0.034}$ \\
\hline
\end{tabular}
     }\eqmoveup
\end{table*}

\begin{figure}[!t]
		\centering
		\includegraphics[width=0.65\textwidth]{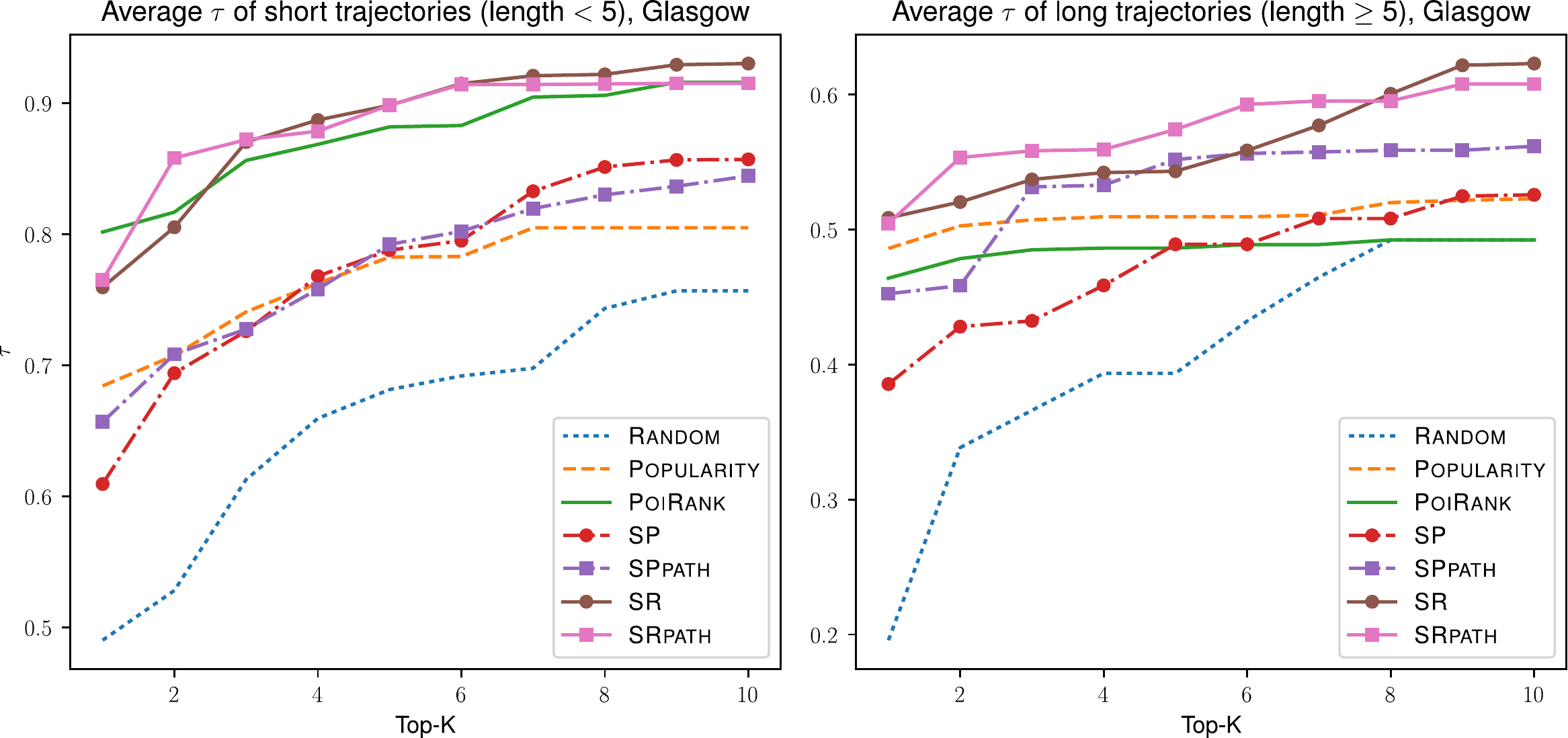}
		\includegraphics[width=0.33\textwidth]{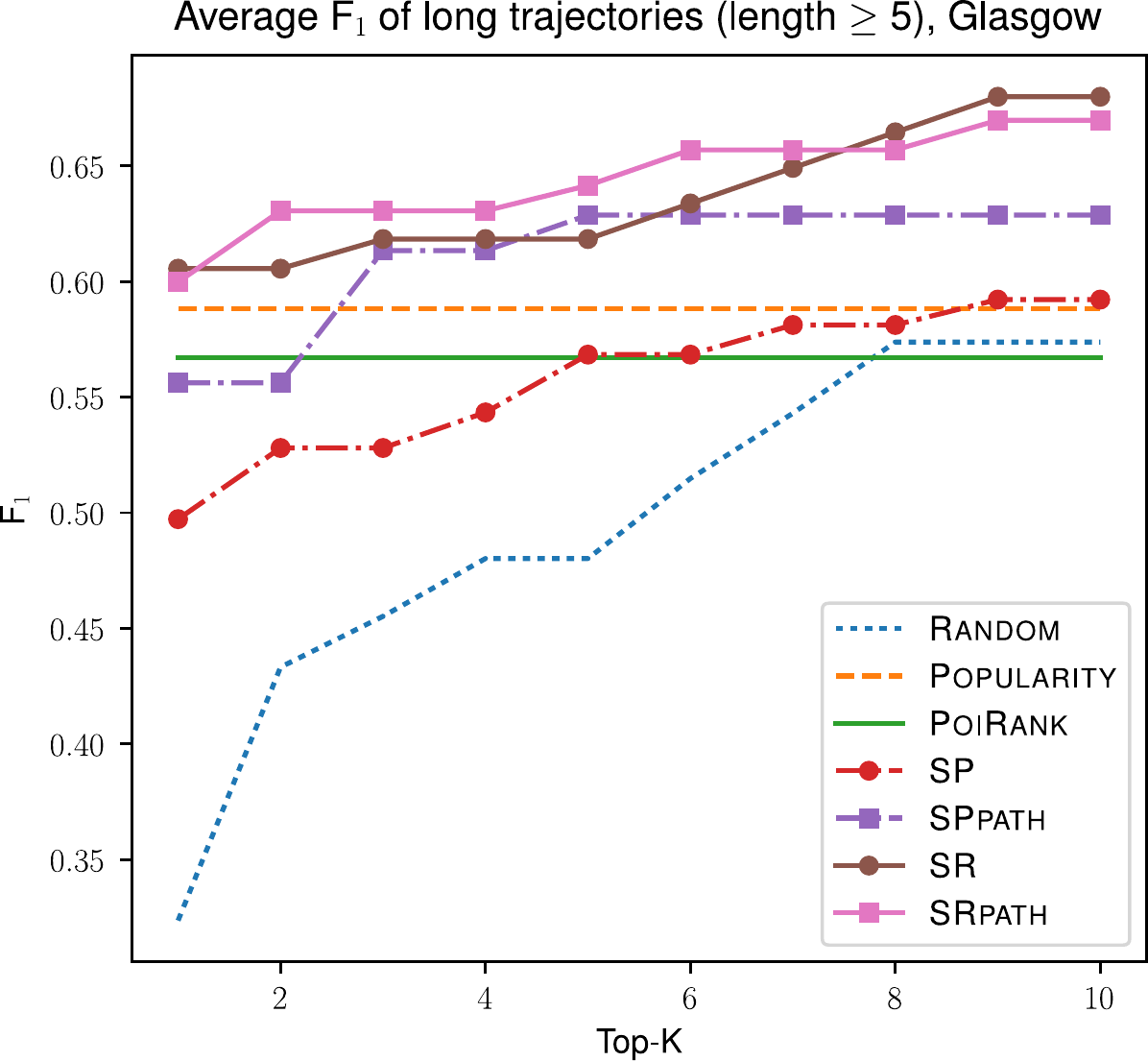}
	    \captionof{figure}{Average Kendall’s $\tau$ over k=1:10, for short (left) and long (middle) trajectories. (right) F$_1$ score on points for long trajectories.
			Structured recommendation methods perform best for all values of $k$.
			For longer trajectories, the predictions of \textsc{Popularity} and \textsc{POIRank} are permutations of the same set of POIs (that do not fully overlap with the ground truth).}
	    \label{fig:topk}
	    \captionmoveup\eqmoveup
\end{figure}


Table~\ref{tab:result} summarises the performance of all methods for top-10 recommendations.
We can observe from the results that \textsc{POIRank} and \textsc{SP},
methods that convert the trajectory recommendation task
into data that is amenable to off the shelf methods (ranking and structured SVM respectively)
performs better than baselines but are not the best performing methods.
The best results are obtained using our proposed methods
\textsc{SPpath}, \textsc{SR}, and \textsc{SRpath}.
We also compare the performance for all values of top-$k$ with $k=1,\ldots,10$, and
Figure~\ref{fig:topk} shows a selection of the curves for Glasgow. We observe that
our proposed methods are consistently the best for all values of $k$.
See the supplement for results across all datasets on all metric variants.
In particular,
accounting for multiple ground truths helps --
\textsc{SR} always performs better than \textsc{SP},
and similarly for the {\sc path} variants of both methods.
This indicates that our first extension -- explicitly modelling multiple ground truths
 -- is important to achieve good performance.
We can also see that the advantages of {\sc SR}, {\sc SPpath}, {\sc SRpath} are salient for longer trajectories, where pairwise and sequential information play a larger role.


Overall, structured recommendation methods have shown superior performance in location sequence recommendation tasks on a public benchmark. Most notably, taking into account multiple ground truths in structured prediction and modeling path constraints are important; and the effects are more pronounced in longer trajectories.
Finally, we note that the unary terms in the sequence scoring function \eqref{eq:jointfeature} can be replaced with {\em personalised} terms to each user, such as from a recommender system~\cite{Koren:2009,bpr09}. We leave this and personalising structured recommendation
as future work.




\secmoveup
\section{Conclusion}
\textmoveup

We cast the problem of sequence recommendation
as a structured prediction problem.
Our proposed solution extends structured SVMs to account for the new setting:
first, we modified the training objective to account for the existence of multiple ground truths;
second, we modified the prediction and loss augmented inference procedures to avoid predicting
loops in the sequence. The inference procedures are novel applications of the list Viterbi
algorithm.
Experiments on real-world trajectory recommendation datasets showed that
structured recommendation outperform existing, non-structured approaches.
Our new viewpoint enables researchers to bring recent advances in structured prediction
to bear on recommender systems problems,
including further improving the efficiency of inference and learning.
In the other direction techniques from recommender systems to capture latent
user- and POI-representations, in sufficiently rich domains, may be used to
improve the predictive power of structured prediction models.

%
\clearpage

\setlength\bibsep{2pt}

\clearpage
\onecolumn

\appendix
{\Large\bf Supplementary material to Structured Recommendation}

\setlength{\belowdisplayskip}{2pt} \setlength{\belowdisplayshortskip}{1pt}
\setlength{\abovedisplayskip}{2pt} \setlength{\abovedisplayshortskip}{1pt}

\section{Model learning and prediction}
\label{sec:supplement}

In this section, we describe the 1-slack formulation for the proposed SR model 
and the details of the list Viterbi algorithm.

\subsection{1-slack formulation for the SR model}
\label{ssec:1slack_sr}

We can use \emph{one} slack variable to represent the sum of the $N$ hinge losses:
\begin{equation*}
\xi_i = \max \left( 0, \, 
        \max_{\bar{\y} \in \mathcal{Y}}
        \left\{ \Delta(\y^{(i)}, \bar{\y}) + \w^\top \Psi(\x^{(i)}, \bar{\y}) \right\} - \w^\top \Psi(\x^{(i)}, \y^{(i)}) \right).
\end{equation*}
Which results in the 1-slack formulation for the SR model:
\begin{equation*}
\resizebox{0.9\linewidth}{!}{$
\begin{aligned}
\min_{\w, \, \xi \ge 0} ~\frac{1}{2} \w^\top \w + C \xi, ~~s.t.~ \frac{1}{N} \left( \sum_{i,j} \w^\top \Psi(\x^{(i)}, \y^{(ij)}) - \w^\top \Psi(\x^{(i)}, \bar{\y}^{(i)}) \right) 
  \ge \frac{1}{N} \sum_{i,j} \Delta(\y^{(ij)}, \bar{\y}^{(i)}) - \xi.
\end{aligned}
$}
\end{equation*}

\subsection{The list Viterbi algorithm}
\label{sec:listviterbi-supp}

We make use of the list Viterbi in four situations:
\begin{enumerate}
  \item To avoid sequence with loops during the prediction phase of the SP and SR models
  \item To make top-$k$ prediction using the SP and SR models
  \item To eliminate known ground truths during the training phase (\ie loss augmented inference) of the SR and \textsc{SRpath} models
  \item To avoid sequence with loops during the training phase (\ie loss augmented inference) of the \textsc{SPpath} and \textsc{SRpath} models
\end{enumerate}


The serial list Viterbi algorithm~\cite{nilsson2001sequentially,seshadri1994list} maintains
a heap (\ie priority queue) of potential solutions, which are then checked for the desired property (for example
whether there are loops). Once the requisite number of trajectories with the desired
property are found, the algorithm terminates (for example once $k$ trajectories without loops are found when performing top-k prediction)
The heap is initialised by running forward-backward (Algorithm~\ref{alg:forward-backward}) followed by the vanilla Viterbi (Algorithm~\ref{alg:viterbi}).

\begin{algorithm}[htbp]
\caption{Forward-backward procedure~\cite{rabiner1989tutorial}}
\label{alg:forward-backward}
\begin{algorithmic}[1]
  \STATE $\forall p_j \in \mathcal{P},~ \alpha_t(p_j) =
          \begin{cases}
          0,~ t = 1 \\
          \max_{p_i \in \mathcal{P}} \left\{ \alpha_{t-1}(p_i) + \mathbf{w}_{ij}^\top \psi_{ij}(\mathbf{x}, p_i, p_j) +
          \mathbf{w}_j^\top \psi_j(\mathbf{x}, p_j) \right\},~ t=2,\dots,l
          \end{cases}$

  \STATE $\forall p_i \in \mathcal{P},~ \beta_t(p_i) =
          \begin{cases}
          0,~ t = l \\
          \max_{p_j \in \mathcal{P}} \left\{ \mathbf{w}_{ij}^\top \psi_{ij}(\mathbf{x}, p_i, p_j) +
          \mathbf{w}_j^\top \psi_j(\mathbf{x}, p_j) + \beta_{t+1}(p_j) \right\},~ t = l-1,\dots,1
          \end{cases}$

  \STATE $\forall p_i, p_j \in \mathcal{P},~ f_{t,t+1}(p_i, p_j) = \alpha_t(p_i) + \mathbf{w}_{ij}^\top \psi_{ij}(\mathbf{x}, p_i, p_j) +
                                \mathbf{w}_j^\top \psi_j(\mathbf{x}, p_j) + \beta_{t+1}(p_j),~ t = 1,\dots,l-1$
\end{algorithmic}
\end{algorithm}

\begin{algorithm}[htbp]
\caption{Viterbi}
\label{alg:viterbi}
\begin{algorithmic}[1]
  \STATE $y_t^1 = \begin{cases}
                  s,~ t = 1 \\
                  \argmax_{p \in \mathcal{P}} \left\{ f_{t-1,t}(y_{t-1}^1, p) \right\},~ t = 2,\dots,l
                  \end{cases}$

  \STATE $r^1 = \max_{p \in \mathcal{P}} \left\{ \alpha_{l}(p) \right\}~~~ \triangleright$ $r^1$ is the score/priority of $\mathbf{y}^1$
\end{algorithmic}
\end{algorithm}

Given an existing heap containing potential trajectories,
list Viterbi maintains a set of POIs $S$ to exclude, which is updated
sequentially by considering the front sequence of the heap.

Recall that for trajectory recommendation we are given the query $\mathbf{x}=(s, l)$, where
$s$ is the starting POI from the set of POIs $\mathcal{P}$,
and $l$ is the desired length of the trajectory.
We assume the score function is of the form $\mathbf{w}^\top \Psi$ where $\Psi$ is the joint
feature vector. $\mathbf{w}$ could be the value of the weight in the current iteration in training,
or the learned weight vector during prediction.

The list Viterbi algorithm for performing top-$k$ prediction is described in Algorithm~\ref{alg:listviterbi}.
To eliminating known ground truths in loss augmented inference,
we modify the forward-backward procedure (Algorithm~\ref{alg:forward-backward}) to account for the loss term $\Delta(\cdot,\cdot)$,
and Algorithm~\ref{alg:viterbi} and Algorithm~\ref{alg:listviterbi} can be used without modification.

\begin{algorithm}[htbp]
\caption{The list Viterbi algorithm for top-$K$ prediction~\cite{nilsson2001sequentially,seshadri1994list}}
\label{alg:listviterbi}
\begin{algorithmic}[1]
\STATE \textbf{Input}: $\mathbf{x}=(s, l),~ \mathcal{P},~ \mathbf{w},~ \Psi, ~K$
\STATE Initialise score matrices $\alpha,~ \beta,~ f_{t, t+1}$, a max-heap $H$, result set $R$, $k=0$.
\STATE $\triangleright$ Do the forward-backward procedure (Algorithm~\ref{alg:forward-backward})
\STATE $\triangleright$ Identify the best (scored) trajectory $\mathbf{y}^1=(y_1^1,\dots,y_l^1)$
  with Viterbi (Algorithm~\ref{alg:viterbi}). This may be a trajectory that violates the desired condition.
\STATE $H.\textit{push}\left(r^1,~ (\mathbf{y}^1, \textsc{nil}, \emptyset) \right)$
\STATE Set $R=\emptyset$, the list of trajectories to be returned.
\WHILE{$H \ne \emptyset$ \textbf{and} $k < \,|\mathcal{P}|^{l-1} - \prod_{t=2}^l (|\mathcal{P}|-t+1) + K$}
    \STATE $r^k,~ (\mathbf{y}^k, I, S) = H.\textit{pop}()~~~ \triangleright$
           $r^k$ is the score of $\mathbf{y}^k=(y_1^k,\dots,y_l^k)$, $I$ is the partition index, and $S$ is the exclude set
    \STATE $k = k + 1$
    \STATE Add $\mathbf{y}^k$ to $R$ if it satisfies the desired property
    \RETURN $R$ if it contains the required number of trajectories
    \STATE $I' = \begin{cases}
                  2, & I = \textsc{nil} \\
                  I, & \text{otherwise}
                 \end{cases}$

    \FOR{$t = I',\dots,l$}
        \STATE $S' = \begin{cases}
                      S \cup \{ y_t^k \}, & t = I' \\
                      \{ y_t^k \},        & \text{otherwise}
                     \end{cases}$

        \STATE $y'_j = \begin{cases}
                            y_j^k,                                                                             & j=1,\dots,t-1 \\
                            \argmax_{p \in \mathcal{P} \setminus S'} \left\{ f_{t-1,t}(y_{t-1}^k, p) \right\}, & j=t \\
                            \argmax_{p \in \mathcal{P}} \left\{ f_{j-1, j}(y'_{j-1}, p) \right\},              & j=t+1,\dots,l
                \end{cases}$
        \STATE \vspace{3pt}$r' = r^k + f_{t-1,t}(y_{t-1}^k, y'_t) - f_{t-1,t}(y_{t-1}^k, y_t^k)$

        \STATE \vspace{3pt}$H.\textit{push}(r', (\y', t, S') )$ \vspace{2pt}
    \ENDFOR
\ENDWHILE
\end{algorithmic}
\end{algorithm}

\clearpage

\section{Experiment}

In this section, we describe trajectory dataset used in experiments as well as features for each methods.
In addition, the details of evaluation and empirical results are also described.

\subsection{Photo trajectory dataset and features}
\label{sec:feature}

\begin{figure}[t]
	\centering
	\includegraphics[width=.7\linewidth]{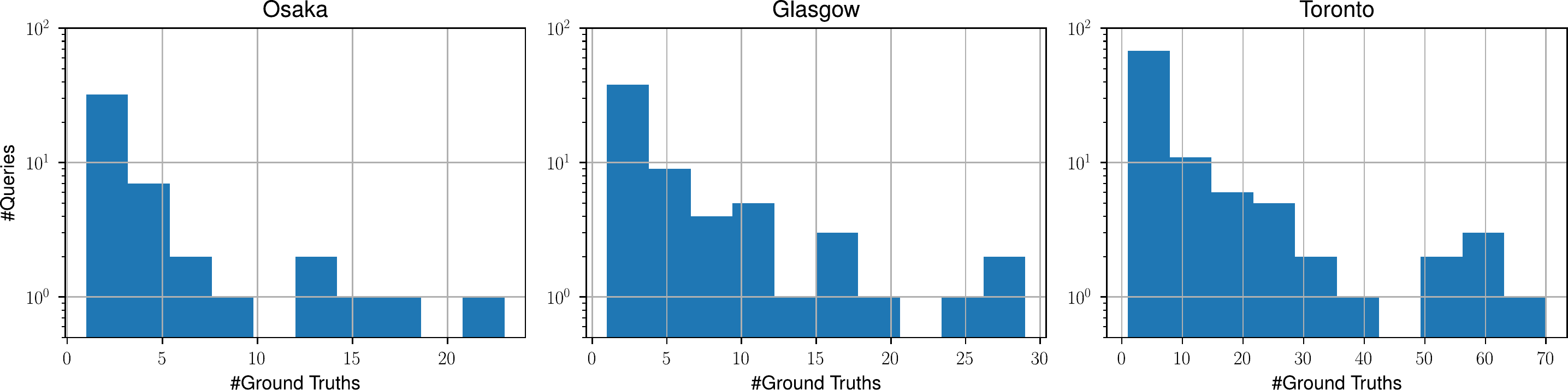}
	\caption{Histograms of the number of trajectories per query.}
	\label{fig:hist_query}
\end{figure}

\begin{figure}[t]
	\centering
	\includegraphics[width=.7\linewidth]{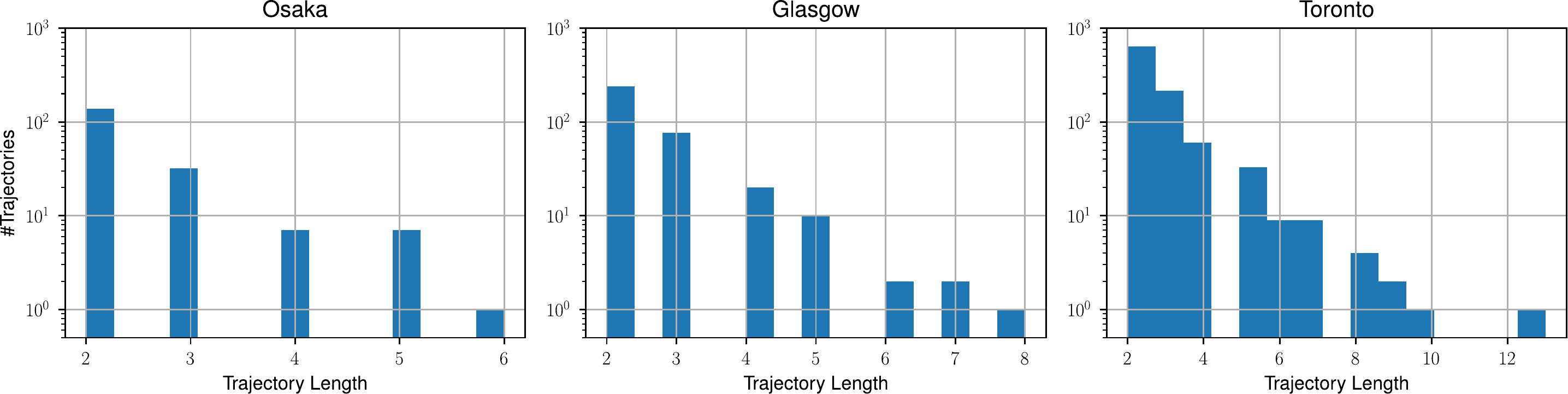}
	\caption{Histograms of trajectory length.}
	\label{fig:hist_length}
\end{figure}

\textbf{Dataset}.
In the interests of reproducibility we present further details of our empirical experiment.
The histogram of the number of trajectories per query is shown in Figure~\ref{fig:hist_query},
where we can see each query has 4-9 ground truths (\ie trajectories) on average, and 30-60 trajectories at most.
The histogram of trajectory length (\ie the number of POIs in a trajectory) is shown in Figure~\ref{fig:hist_length},
where we can see the majority are short trajectories (\ie length $\le$ 5).

\textbf{Features}.
The POI-query features used by \textsc{PoiRank}, SP and SR methods and their extensions 
(\ie the \textsc{SPpath} and \textsc{SRpath} models) are shown in Table~\ref{tab:poifeature},
pairwise features used in SP and SR methods and their extensions are shown in Table~\ref{tab:tranfeature}.

\begin{table*}[!h]
\caption{POI-query features: features of POI $p$ with respect to query $(s,l)$}
\label{tab:poifeature}
\centering
\small
\setlength{\tabcolsep}{10pt} 
\begin{tabular}{l|l} \hline
\textbf{Feature}       & \textbf{Description} \\ \hline
\texttt{category}      & one-hot encoding of the category of $p$ \\
\texttt{neighbourhood} & one-hot encoding of the POI cluster that $p$ resides in \\
\texttt{popularity}    & logarithm of POI popularity of $p$ \\
\texttt{nVisit}        & logarithm of the total number of visit by all users at $p$ \\
\texttt{avgDuration}  & logarithm of the average visit duration at $p$ \\
\hline

\texttt{trajLen}           & trajectory length $l$, i.e., the number of POIs required \\
\texttt{sameCatStart}      & $1$ if the category of $p$ is the same as that of $s$, $-1$ otherwise \\
\texttt{sameNeighbourhoodStart} & $1$ if $p$ resides in the same POI cluster as $s$, $-1$ otherwise \\
\texttt{diffPopStart}    & real-valued difference in POI popularity of $p$ from that of $s$ \\
\texttt{diffNVisitStart}        & real-valued difference in the total number of visit at $p$ from that at $s$ \\
\texttt{diffDurationStart}  & real-valued difference in average duration at $p$ from that at $s$ \\
\texttt{distStart}          & distance between $p$ and $s$, calculated using the Haversine formula \\
\hline
\end{tabular}
\end{table*}

\begin{table}[!h]
\caption{Pairwise POI features}
\label{tab:tranfeature}
\centering
\small
\setlength{\tabcolsep}{2pt} 
\begin{tabular}{l|l} \hline
\textbf{Feature}       & \textbf{Description} \\ \hline
\texttt{category}      & category of POI \\
\texttt{neighbourhood} & the cluster that a POI resides in \\
\texttt{popularity}    & (discretised) popularity of POI \\
\texttt{nVisit}        & (discretised) total number of visit at POI \\
\texttt{avgDuration}  & (discretised) average duration at POI \\ \hline
\end{tabular}
\end{table}

\clearpage
\subsection{Evaluation settings}
\label{sec:metric}

\textbf{Top-k prediction for baselines}.
\begin{itemize}
\item To perform top-$k$ prediction with \textsc{Random} baseline, we simply repeat the \textsc{Random} method $k$ times.
\item To perform top-$k$ prediction with \textsc{Popularity} and \textsc{PoiRank}, we make use of the list Viterbi algorithm 
      (Algorithm~\ref{alg:listviterbi} to get $k$ best scored paths, in particular, 
      for \textsc{Popularity}, the score of a path is the accumulated popularity of all POIs in the path; 
      for \textsc{PoiRank}, the score of a path is the likelihood 
      (the ranking scores for POIs are first transformed into a probability distribution using the softmax function, as described in~\cite{cikm16paper}).
\end{itemize}

To evaluate the performance of a certain recommendation algorithm,
we need to measure the similarity (or loss) given prediction $\hat{\mathbf{y}}$
and ground truth $\mathbf{y}$.

\textbf{F$_1$ score on points}.
F$_1$ score on points~\cite{ijcai15} cares about only the set of correctly recommended POIs.
\begin{equation*}
F_1(\mathbf{y}, \hat{\mathbf{y}}) = \frac{2  P_{\textsc{point}}  R_{\textsc{point}}}{P_{\textsc{point}} + R_{\textsc{point}}}
\end{equation*}
where $P_\textsc{point}$, $R_\textsc{point}$ are respectively the precision and recall for points in $\hat\y$ and $\y$.
If $| \hat{\mathbf{y}} | = | \mathbf{y} |$, this metric is just the unordered Hamming loss,
i.e., Hamming loss between two binary indicator vectors of size $| \mathcal{P} |$.

\textbf{F$_1$ score on pairs}.
To take into account the orders in recommended sequence, 
we also use the F$_1$ score on pairs~\cite{cikm16paper} measure, which considers the set of correctly predicted POI pairs,
\begin{equation*}
\text{pairs-F}_1(\mathbf{y}, \hat{\mathbf{y}}) = \frac{2 P_{\textsc{pair}} R_{\textsc{pair}}}{P_{\textsc{pair}} + R_{\textsc{pair}}}
\end{equation*}
where $P_\textsc{point}$, $R_\textsc{point}$ are respectively the precision and recall for all possible pairs of $\hat\y$ and $\y$.

\textbf{Kendall's $\tau$ with ties}
Alternatively, we can cast a trajectory $\y = y_{1:l}$ as a ranking of POIs in $\mathcal{P}$,
where $y_j$ has a rank $| \mathcal{P} | - j + 1$ and any other POI $p \notin \mathbf{y}$ has a rank $0$ ($0$ is an arbitrary choice),
then we can make use of ranking evaluation metrics such as Kendall's $\tau$ by taking care of ties in ranks.
In particular, given a prediction $\hat\y = \hat{y_{1:l}}$ and ground truth $\y = y_{1:l}$,
we produce two ranks for $\mathbf{y}$ and $\hat{\mathbf{y}}$ with respect to 
a specific ordering of POIs $(p_1, p_2, \dots, p_{|\mathcal{P}|})$:
\begin{align*}
r_i       &= \sum_{j=1}^l (| \mathcal{P} | - j + 1)  \llb p_i = y_j \rrb,~
i = 1, \dots, | \mathcal{P} | \\
\hat{r}_i &= \sum_{j=1}^l (| \mathcal{P} | - j + 1)  \llb p_i = \hat{y}_j \rrb,~
i = 1, \dots, | \mathcal{P} |
\end{align*}
where POIs not in $\mathbf{y}$ will have a rank of $0$.
Then we compute the following metrics:
\begin{itemize}
\item the number of concordant pairs \(
      C = \frac{1}{2} \sum_{i,j} \left(\llb r_i < r_j \rrb  \llb \hat{r}_i < \hat{r}_j \rrb +
                      \llb r_i > r_j \rrb  \llb \hat{r}_i > \hat{r}_j \rrb \right) \)
\item the number of discordant pairs \(
      D = \frac{1}{2} \sum_{i,j} \left(\llb r_i < r_j \rrb  \llb \hat{r}_i > \hat{r}_j \rrb +
                      \llb r_i > r_j \rrb  \llb \hat{r}_i < \hat{r}_j \rrb \right) \)
\item the number of ties in ground truth $\y$: \(
      T_{\mathbf{y}} = \frac{1}{2} \sum_{i \ne j} \llb r_i = r_j \rrb 
                     = \frac{1}{2} \left( |\mathcal{P}| - l \right) \left( |\mathcal{P}| - l - 1 \right) \)
\item the number of ties in prediction $\hat\y$: \(
      T_{\hat{\mathbf{y}}} = \frac{1}{2} \sum_{i \ne j} \llb \hat{r}_i = \hat{r}_j \rrb 
                           = \frac{1}{2} \left( |\mathcal{P}| - l \right) \left( |\mathcal{P}| - l - 1 \right) \)
\item the number of ties in both $\y$ and $\hat\y$: \(
      T_{\mathbf{y},\hat{\mathbf{y}}} = \frac{1}{2} \sum_{i \ne j} \llb r_i = r_j \rrb  \llb \hat{r}_i = \hat{r}_j \rrb \)
\end{itemize}
Kendall's $\tau$ (version $b$)~\cite{kendall1945,agresti2010analysis} is
\begin{equation*}
\tau_b(\mathbf{y}, \hat{\mathbf{y}}) = \frac{C - D}{\sqrt{(C + D + T) (C + D + U)}},
\end{equation*}
where $T = T_{\mathbf{y}} - T_{\mathbf{y},\hat{\mathbf{y}}}$ and $U = T_{\hat{\mathbf{y}}} - T_{\mathbf{y},\hat{\mathbf{y}}}$.


\clearpage
\subsection{Empirical results}

\textbf{The effects of $k$ for top-$k$ prediction}.
The performance of baselines and structured recommendation algorithms for top-$k$ ($k=1,3,5,10$) 
prediction are shown in the following tables.
Bold entries: \textbf{best} performing method for each metric; italicised entries: the \textit{next best}. 

\begin{table*}[!h]
\caption{Results on trajectory recommendation datasets on best of top-1.}
\centering
\scriptsize
\setlength{\tabcolsep}{3pt} 
\begin{tabular}{l|cc|cc|ccc} \hline
& \multicolumn{7}{c}{\bf Kendall's $\tau$} \\ \hline
 & \textsc{Random} & \textsc{Popularity} & \textsc{PoiRank} & \textsc{SP} & \textsc{SPpath} & \textsc{SR} & \textsc{SRpath} \\ \hline
Glasgow & $0.430\pm0.031$ & $0.644\pm0.036$ & $\mathbf{0.733\pm0.030}$ & $0.564\pm0.029$ & $0.615\pm0.034$ & $0.708\pm0.031$ & $\mathit{0.712\pm0.031}$ \\
Osaka & $0.420\pm0.030$ & $0.566\pm0.034$ & $\mathbf{0.644\pm0.040}$ & $0.525\pm0.037$ & $0.525\pm0.039$ & $0.608\pm0.042$ & $\mathit{0.613\pm0.044}$ \\
Toronto & $0.394\pm0.025$ & $0.626\pm0.023$ & $\mathit{0.714\pm0.024}$ & $0.543\pm0.026$ & $0.572\pm0.026$ & $0.714\pm0.026$ & $\mathbf{0.717\pm0.026}$ \\
\hline
& \multicolumn{7}{c}{\bf F$_1$ score on points} \\ \hline
Glasgow & $0.478\pm0.027$ & $0.681\pm0.032$ & $\mathbf{0.764\pm0.027}$ & $0.604\pm0.026$ & $0.653\pm0.031$ & $0.741\pm0.028$ & $\mathit{0.743\pm0.028}$ \\
Osaka & $0.459\pm0.027$ & $0.601\pm0.031$ & $\mathbf{0.678\pm0.037}$ & $0.555\pm0.034$ & $0.558\pm0.036$ & $0.638\pm0.039$ & $\mathit{0.645\pm0.040}$ \\
Toronto & $0.461\pm0.020$ & $0.671\pm0.021$ & $\mathit{0.756\pm0.021}$ & $0.594\pm0.023$ & $0.623\pm0.023$ & $0.753\pm0.023$ & $\mathbf{0.757\pm0.022}$ \\
\hline
& \multicolumn{7}{c}{\bf F$_1$ score on pairs} \\ \hline
Glasgow & $0.154\pm0.035$ & $0.426\pm0.051$ & $\mathbf{0.545\pm0.046}$ & $0.289\pm0.042$ & $0.389\pm0.048$ & $0.506\pm0.048$ & $\mathit{0.516\pm0.048}$ \\
Osaka & $0.104\pm0.037$ & $0.281\pm0.051$ & $\mathbf{0.428\pm0.059}$ & $0.243\pm0.052$ & $0.254\pm0.055$ & $0.375\pm0.059$ & $\mathit{0.401\pm0.060}$ \\
Toronto & $0.143\pm0.025$ & $0.384\pm0.034$ & $0.506\pm0.036$ & $0.299\pm0.033$ & $0.340\pm0.035$ & $\mathit{0.530\pm0.037}$ & $\mathbf{0.533\pm0.037}$ \\
\hline
\end{tabular}
\end{table*}

\begin{table*}[!h]
\caption{Results on trajectory recommendation datasets on best of top-3.}
\centering
\scriptsize
\setlength{\tabcolsep}{3pt} 
\begin{tabular}{l|cc|cc|ccc} \hline
& \multicolumn{7}{c}{\bf Kendall's $\tau$} \\ \hline
 & \textsc{Random} & \textsc{Popularity} & \textsc{PoiRank} & \textsc{SP} & \textsc{SPpath} & \textsc{SR} & \textsc{SRpath} \\ \hline
Glasgow & $0.563\pm0.031$ & $0.693\pm0.036$ & $0.781\pm0.030$ & $0.666\pm0.033$ & $0.688\pm0.032$ & $\mathit{0.803\pm0.029}$ & $\mathbf{0.808\pm0.030}$ \\
Osaka & $0.556\pm0.037$ & $0.666\pm0.039$ & $\mathbf{0.726\pm0.042}$ & $0.630\pm0.044$ & $0.698\pm0.040$ & $\mathit{0.711\pm0.042}$ & $0.697\pm0.042$ \\
Toronto & $0.521\pm0.026$ & $0.670\pm0.025$ & $0.746\pm0.023$ & $0.629\pm0.027$ & $0.650\pm0.027$ & $\mathbf{0.753\pm0.025}$ & $\mathit{0.749\pm0.024}$ \\
\hline
& \multicolumn{7}{c}{\bf F$_1$ score on points} \\ \hline
Glasgow & $0.598\pm0.028$ & $0.722\pm0.033$ & $0.803\pm0.027$ & $0.698\pm0.030$ & $0.716\pm0.029$ & $\mathit{0.825\pm0.026}$ & $\mathbf{0.829\pm0.026}$ \\
Osaka & $0.587\pm0.034$ & $0.691\pm0.035$ & $\mathbf{0.750\pm0.039}$ & $0.656\pm0.040$ & $0.724\pm0.037$ & $\mathit{0.735\pm0.038}$ & $0.723\pm0.039$ \\
Toronto & $0.577\pm0.022$ & $0.704\pm0.023$ & $0.776\pm0.021$ & $0.674\pm0.023$ & $0.693\pm0.023$ & $\mathbf{0.784\pm0.022}$ & $\mathit{0.780\pm0.021}$ \\
\hline
& \multicolumn{7}{c}{\bf F$_1$ score on pairs} \\ \hline
Glasgow & $0.300\pm0.043$ & $0.524\pm0.053$ & $0.625\pm0.046$ & $0.464\pm0.049$ & $0.481\pm0.048$ & $\mathit{0.666\pm0.045}$ & $\mathbf{0.678\pm0.045}$ \\
Osaka & $0.288\pm0.055$ & $0.448\pm0.058$ & $\mathbf{0.578\pm0.060}$ & $0.425\pm0.062$ & $0.511\pm0.059$ & $\mathit{0.549\pm0.060}$ & $0.520\pm0.059$ \\
Toronto & $0.281\pm0.032$ & $0.477\pm0.036$ & $0.575\pm0.035$ & $0.429\pm0.037$ & $0.461\pm0.037$ & $\mathbf{0.592\pm0.036}$ & $\mathit{0.584\pm0.036}$ \\
\hline
\end{tabular}
\end{table*}

\begin{table*}[!h]
\caption{Results on trajectory recommendation datasets on best of top-5.}
\centering
\scriptsize
\setlength{\tabcolsep}{3pt} 
\begin{tabular}{l|cc|cc|ccc} \hline
& \multicolumn{7}{c}{\bf Kendall's $\tau$} \\ \hline
 & \textsc{Random} & \textsc{Popularity} & \textsc{PoiRank} & \textsc{SP} & \textsc{SPpath} & \textsc{SR} & \textsc{SRpath} \\ \hline
Glasgow & $0.623\pm0.029$ & $0.727\pm0.037$ & $0.801\pm0.030$ & $0.727\pm0.033$ & $0.743\pm0.031$ & $\mathit{0.826\pm0.028}$ & $\mathbf{0.832\pm0.028}$ \\
Osaka & $0.618\pm0.038$ & $0.674\pm0.038$ & $\mathbf{0.750\pm0.040}$ & $0.678\pm0.045$ & $0.735\pm0.039$ & $\mathit{0.741\pm0.039}$ & $0.729\pm0.041$ \\
Toronto & $0.574\pm0.025$ & $0.687\pm0.025$ & $0.754\pm0.023$ & $0.662\pm0.027$ & $0.683\pm0.026$ & $\mathbf{0.778\pm0.023}$ & $\mathit{0.769\pm0.024}$ \\
\hline
& \multicolumn{7}{c}{\bf F$_1$ score on points} \\ \hline
Glasgow & $0.655\pm0.026$ & $0.754\pm0.033$ & $0.821\pm0.026$ & $0.755\pm0.030$ & $0.770\pm0.027$ & $\mathit{0.847\pm0.024}$ & $\mathbf{0.850\pm0.025}$ \\
Osaka & $0.646\pm0.035$ & $0.699\pm0.034$ & $\mathbf{0.772\pm0.037}$ & $0.700\pm0.041$ & $0.757\pm0.036$ & $\mathit{0.761\pm0.036}$ & $0.751\pm0.037$ \\
Toronto & $0.624\pm0.022$ & $0.719\pm0.023$ & $0.781\pm0.021$ & $0.705\pm0.023$ & $0.724\pm0.022$ & $\mathbf{0.808\pm0.021}$ & $\mathit{0.798\pm0.021}$ \\
\hline
& \multicolumn{7}{c}{\bf F$_1$ score on pairs} \\ \hline
Glasgow & $0.377\pm0.044$ & $0.590\pm0.052$ & $0.670\pm0.045$ & $0.563\pm0.048$ & $0.573\pm0.047$ & $\mathit{0.701\pm0.043}$ & $\mathbf{0.715\pm0.044}$ \\
Osaka & $0.375\pm0.058$ & $0.459\pm0.057$ & $\mathbf{0.607\pm0.058}$ & $0.507\pm0.064$ & $0.568\pm0.058$ & $\mathit{0.584\pm0.058}$ & $0.575\pm0.058$ \\
Toronto & $0.343\pm0.034$ & $0.504\pm0.036$ & $0.593\pm0.034$ & $0.483\pm0.037$ & $0.510\pm0.037$ & $\mathbf{0.624\pm0.035}$ & $\mathit{0.610\pm0.035}$ \\
\hline
\end{tabular}
\end{table*}

\begin{table*}[!h]
\caption{Results on trajectory recommendation datasets on best of top-10.}
\centering
\scriptsize
\setlength{\tabcolsep}{3pt} 
\begin{tabular}{l|cc|cc|ccc} \hline
& \multicolumn{7}{c}{\bf Kendall's $\tau$} \\ \hline
 & \textsc{Random} & \textsc{Popularity} & \textsc{PoiRank} & \textsc{SP} & \textsc{SPpath} & \textsc{SR} & \textsc{SRpath} \\ \hline
Glasgow & $0.703\pm0.029$ & $0.748\pm0.036$ & $0.830\pm0.029$ & $0.790\pm0.030$ & $0.787\pm0.029$ & $\mathbf{0.868\pm0.026}$ & $\mathit{0.853\pm0.026}$ \\
Osaka & $0.685\pm0.035$ & $0.768\pm0.038$ & $0.787\pm0.037$ & $0.749\pm0.043$ & $\mathit{0.791\pm0.036}$ & $0.777\pm0.036$ & $\mathbf{0.803\pm0.034}$ \\
Toronto & $0.652\pm0.024$ & $0.719\pm0.024$ & $0.784\pm0.023$ & $0.697\pm0.027$ & $0.719\pm0.026$ & $\mathbf{0.802\pm0.022}$ & $\mathit{0.797\pm0.022}$ \\
\hline
& \multicolumn{7}{c}{\bf F$_1$ score on points} \\ \hline
Glasgow & $0.731\pm0.026$ & $0.771\pm0.033$ & $0.847\pm0.025$ & $0.810\pm0.027$ & $0.807\pm0.026$ & $\mathbf{0.883\pm0.023}$ & $\mathit{0.868\pm0.023}$ \\
Osaka & $0.703\pm0.032$ & $0.786\pm0.034$ & $0.804\pm0.034$ & $0.770\pm0.039$ & $\mathit{0.809\pm0.033}$ & $0.793\pm0.033$ & $\mathbf{0.820\pm0.031}$ \\
Toronto & $0.696\pm0.021$ & $0.746\pm0.022$ & $0.807\pm0.020$ & $0.733\pm0.023$ & $0.755\pm0.022$ & $\mathbf{0.828\pm0.019}$ & $\mathit{0.823\pm0.020}$ \\
\hline
& \multicolumn{7}{c}{\bf F$_1$ score on pairs} \\ \hline
Glasgow & $0.495\pm0.046$ & $0.623\pm0.051$ & $0.726\pm0.043$ & $0.658\pm0.046$ & $0.648\pm0.045$ & $\mathbf{0.770\pm0.039}$ & $\mathit{0.746\pm0.041}$ \\
Osaka & $0.451\pm0.057$ & $0.626\pm0.055$ & $0.661\pm0.056$ & $0.620\pm0.061$ & $\mathit{0.664\pm0.055}$ & $0.637\pm0.055$ & $\mathbf{0.671\pm0.053}$ \\
Toronto & $0.438\pm0.034$ & $0.550\pm0.035$ & $0.649\pm0.033$ & $0.530\pm0.037$ & $0.552\pm0.036$ & $\mathbf{0.660\pm0.033}$ & $\mathit{0.657\pm0.034}$ \\
\hline
\end{tabular}
\end{table*}

\textbf{Performance on short and long trajectories}.
The performance of baselines and structured recommendation algorithms 
for short (length $<$ 5) and long (length $\ge$ 5) trajectories 
with top-$k$ ($k=1:10$) prediction are shown in the following figures.

\includepdf[pages={1-}]{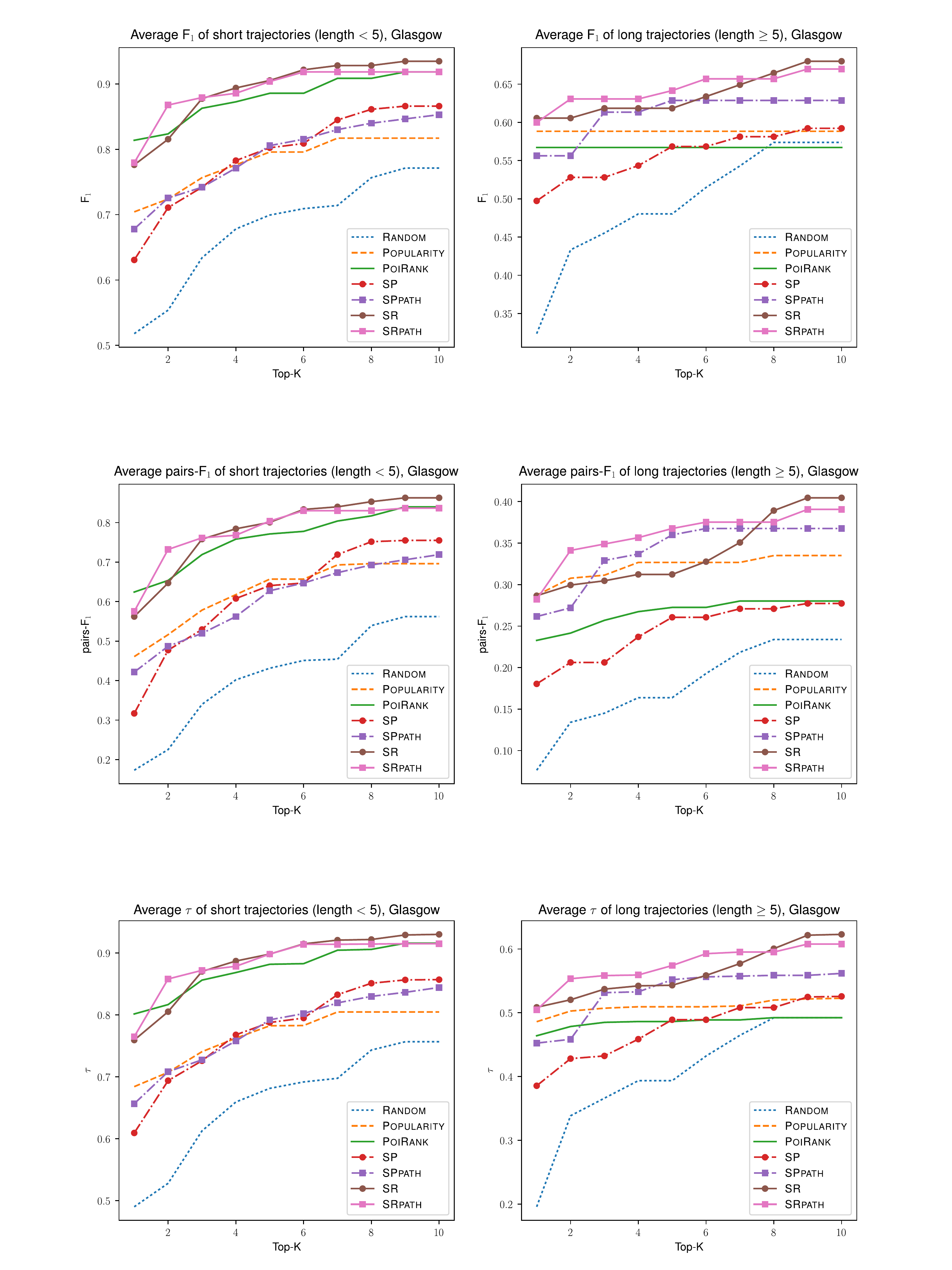}


\begin{thebibliography}{10}

\bibitem{Goldberg:1992}
D.~Goldberg, D.~Nichols, B.~M. Oki, and D.~Terry, ``Using collaborative
  filtering to weave an information tapestry,'' {\em Commun. ACM}, vol.~35,
  Dec. 1992.

\bibitem{Sarwar:2001}
B.~Sarwar, G.~Karypis, J.~Konstan, and J.~Riedl, ``Item-based collaborative
  filtering recommendation algorithms,'' in {\em WWW '01}, (New York, NY, USA),
  pp.~285--295, ACM, 2001.

\bibitem{Koren:2010}
Y.~Koren, ``{Factor in the neighbors: Scalable and accurate collaborative
  filtering},'' {\em ACM Trans. Knowl. Discov. Data}, vol.~4, pp.~1:1----1:24,
  Jan. 2010.

\bibitem{Linden:2003}
G.~Linden, B.~Smith, and J.~York, ``{Amazon.com Recommendations: Item-to-Item
  Collaborative Filtering},'' {\em IEEE Internet Computing}, vol.~7,
  pp.~76--80, Jan. 2003.

\bibitem{Agarwal:2013}
D.~Agarwal, B.-C. Chen, P.~Elango, and R.~Ramakrishnan, ``Content
  recommendation on web portals,'' {\em Commun. ACM}, vol.~56, pp.~92--101,
  June 2013.

\bibitem{Amatriain:2015}
X.~Amatriain and J.~Basilico, {\em Recommender Systems in Industry: A {Netflix}
  Case Study}, pp.~385--419.
\newblock Boston, MA: Springer US, 2015.

\bibitem{Gomez-Uribe:2015}
C.~A. Gomez-Uribe and N.~Hunt, ``The {Netflix} recommender system: Algorithms,
  business value, and innovation,'' {\em ACM Transactions on Management
  Information Systems}, vol.~6, pp.~13:1--13:19, Dec. 2015.

\bibitem{Koren:2009}
Y.~Koren, R.~Bell, and C.~Volinsky, ``Matrix factorization techniques for
  recommender systems,'' {\em Computer}, vol.~42, pp.~30--37, Aug. 2009.

\bibitem{lu2010photo2trip}
X.~Lu, C.~Wang, J.-M. Yang, Y.~Pang, and L.~Zhang, ``{Photo2Trip}: Generating
  travel routes from geo-tagged photos for trip planning,'' MM '10,
  pp.~143--152, ACM, 2010.

\bibitem{lu2012personalized}
E.~H.-C. Lu, C.-Y. Chen, and V.~S. Tseng, ``Personalized trip recommendation
  with multiple constraints by mining user check-in behaviors,'' SIGSPATIAL
  '12, pp.~209--218, ACM, 2012.

\bibitem{ijcai15}
K.~H. Lim, J.~Chan, C.~Leckie, and S.~Karunasekera, ``Personalized tour
  recommendation based on user interests and points of interest visit
  durations,'' in {\em Proceedings of the 24th International Joint Conference
  on Artificial Intelligence}, pp.~1778--1784, 2015.

\bibitem{cikm16paper}
D.~Chen, C.~S. Ong, and L.~Xie, ``Learning points and routes to recommend
  trajectories,'' in {\em Proceedings of the 25th ACM International Conference
  on Information and Knowledge Management}, pp.~2227--2232, 2016.

\bibitem{McFee:2011}
B.~McFee and G.~Lanckriet, ``The natural language of playlists,'' in {\em 12th
  International Symposium for Music Information Retrieval (ISMIR2011))},
  October 2011.

\bibitem{chen2012playlist}
S.~Chen, J.~L. Moore, D.~Turnbull, and T.~Joachims, ``Playlist prediction via
  metric embedding,'' in {\em Proceedings of the 18th ACM SIGKDD international
  conference on Knowledge discovery and data mining}, pp.~714--722, ACM, 2012.

\bibitem{hidasi2015session}
B.~Hidasi, A.~Karatzoglou, L.~Baltrunas, and D.~Tikk, ``Session-based
  recommendations with recurrent neural networks,'' {\em arXiv preprint
  arXiv:1511.06939}, 2015.

\bibitem{choi2016towards}
K.~Choi, G.~Fazekas, and M.~Sandler, ``Towards playlist generation algorithms
  using {RNN}s trained on within-track transitions,'' {\em arXiv preprint
  arXiv:1606.02096}, 2016.

\bibitem{dehaspe1998finding}
L.~Dehaspe, H.~Toivonen, and R.~D. King, ``Finding frequent substructures in
  chemical compounds.,'' in {\em KDD}, vol.~98, p.~1998, 1998.

\bibitem{antikacioglu2015recommendation}
A.~Antikacioglu, R.~Ravi, and S.~Sridhar, ``Recommendation subgraphs for web
  discovery,'' in {\em Proceedings of the 24th International Conference on
  World Wide Web}, pp.~77--87, ACM, 2015.

\bibitem{tsochantaridis2005large}
I.~Tsochantaridis, T.~Joachims, T.~Hofmann, and Y.~Altun, ``Large margin
  methods for structured and interdependent output variables,'' {\em Journal of
  Machine Learning Research}, vol.~6, no.~Sep, pp.~1453--1484, 2005.

\bibitem{joachims2009predicting}
T.~Joachims, T.~Hofmann, Y.~Yue, and C.-N. Yu, ``Predicting structured objects
  with support vector machines,'' {\em Communications of the ACM}, vol.~52,
  no.~11, pp.~97--104, 2009.

\bibitem{west2001introduction}
D.~B. West {\em et~al.}, {\em Introduction to graph theory}, vol.~2.
\newblock Prentice hall Upper Saddle River, 2001.

\bibitem{Netflix}
Netflix, ``{Netflix Prize}.'' http://www.netflixprize.com/, 2006.

\bibitem{bao2015recommendations}
J.~Bao, Y.~Zheng, D.~Wilkie, and M.~Mokbel, ``Recommendations in location-based
  social networks: a survey,'' {\em GeoInformatica}, vol.~19, no.~3,
  pp.~525--565, 2015.

\bibitem{zheng2015trajectory}
Y.~Zheng, ``Trajectory data mining: an overview,'' {\em ACM Transactions on
  Intelligent Systems and Technology}, vol.~6, no.~3, p.~29, 2015.

\bibitem{zheng2014urban}
Y.~Zheng, L.~Capra, O.~Wolfson, and H.~Yang, ``Urban computing: concepts,
  methodologies, and applications,'' {\em ACM Transactions on Intelligent
  Systems and Technology}, vol.~5, no.~3, p.~38, 2014.

\bibitem{Rendle:2010}
S.~Rendle, C.~Freudenthaler, and L.~Schmidt-Thieme, ``Factorizing personalized
  markov chains for next-basket recommendation,'' in {\em WWW '10}, (New York,
  NY, USA), pp.~811--820, ACM, 2010.

\bibitem{Wang:2015}
P.~Wang, J.~Guo, Y.~Lan, J.~Xu, S.~Wan, and X.~Cheng, ``Learning hierarchical
  representation model for nextbasket recommendation,'' in {\em SIGIR '15},
  (New York, NY, USA), pp.~403--412, ACM, 2015.

\bibitem{yu2016dynamic}
F.~Yu, Q.~Liu, S.~Wu, L.~Wang, and T.~Tan, ``A dynamic recurrent model for next
  basket recommendation,'' in {\em Proceedings of the 39th International ACM
  SIGIR conference on Research and Development in Information Retrieval},
  pp.~729--732, ACM, 2016.

\bibitem{bpr09}
S.~Rendle, C.~Freudenthaler, Z.~Gantner, and L.~Schmidt-Thieme, ``{BPR}:
  {Bayesian} personalized ranking from implicit feedback,'' UAI '09,
  pp.~452--461, AUAI Press, 2009.

\bibitem{ratliff2006subgradient}
N.~Ratliff, J.~A. Bagnell, and M.~Zinkevich, ``Subgradient methods for maximum
  margin structured learning,'' in {\em ICML workshop on learning in structured
  output spaces}, vol.~46, Citeseer, 2006.

\bibitem{lacoste2013block}
S.~Lacoste-julien, M.~Jaggi, M.~Schmidt, and P.~Pletscher, ``Block-coordinate
  {Frank-Wolfe} optimization for structural {SVM}s,'' in {\em Proceedings of
  the 30th International Conference on Machine Learning (ICML'13)}, pp.~53--61,
  2013.

\bibitem{seshadri1994list}
N.~Seshadri and C.-E. Sundberg, ``List {Viterbi} decoding algorithms with
  applications,'' {\em IEEE Transactions on Communications}, vol.~42, no.~234,
  pp.~313--323, 1994.

\bibitem{nill1995list}
C.~Nill and C.-E. Sundberg, ``List and soft symbol output viterbi algorithms:
  Extensions and comparisons,'' {\em IEEE Transactions on Communications},
  vol.~43, no.~234, pp.~277--287, 1995.

\bibitem{soong1991tree}
F.~K. Soong and E.-F. Huang, ``A tree-trellis based fast search for finding the
  n-best sentence hypotheses in continuous speech recognition,'' in {\em
  Acoustics, Speech, and Signal Processing, 1991. ICASSP-91., 1991
  International Conference on}, pp.~705--708, IEEE, 1991.

\bibitem{nilsson2001sequentially}
D.~Nilsson and J.~Goldberger, ``Sequentially finding the {N}-best list in
  hidden {Markov} models,'' in {\em Proceedings of the 17th international joint
  conference on Artificial intelligence}, pp.~1280--1285, Morgan Kaufmann
  Publishers Inc., 2001.

\bibitem{tspbook2011}
D.~L. Applegate, R.~E. Bixby, V.~Chvatal, and W.~J. Cook, {\em The traveling
  salesman problem: a computational study}.
\newblock Princeton university press, 2011.

\bibitem{lian2014geomf}
D.~Lian, C.~Zhao, X.~Xie, G.~Sun, E.~Chen, and Y.~Rui, ``{GeoMF}: Joint
  geographical modeling and matrix factorization for point-of-interest
  recommendation,'' KDD '14, pp.~831--840, ACM, 2014.

\bibitem{christofides1976}
N.~Christofides, ``Worst-case analysis of a new heuristic for the travelling
  salesman problem,'' Tech. Rep. 388, Graduate School of Industrial
  Administration, CMU, 1976.

\bibitem{thomee2016yfcc100m}
B.~Thomee, B.~Elizalde, D.~A. Shamma, K.~Ni, G.~Friedland, D.~Poland, D.~Borth,
  and L.-J. Li, ``{YFCC100M}: The new data in multimedia research,'' {\em
  Communications of the ACM}, vol.~59, no.~2, pp.~64--73, 2016.

\bibitem{burman1989comparative}
P.~Burman, ``A comparative study of ordinary cross-validation, v-fold
  cross-validation and the repeated learning-testing methods,'' {\em
  Biometrika}, vol.~76, no.~3, pp.~503--514, 1989.

\bibitem{agresti2010analysis}
A.~Agresti, {\em Analysis of ordinal categorical data}, vol.~656.
\newblock John Wiley \& Sons, 2010.

\bibitem{russakovsky2015imagenet}
O.~Russakovsky, J.~Deng, H.~Su, J.~Krause, S.~Satheesh, S.~Ma, Z.~Huang,
  A.~Karpathy, A.~Khosla, M.~Bernstein, {\em et~al.}, ``{ImageNet} large scale
  visual recognition challenge,'' {\em International Journal of Computer
  Vision}, vol.~115, no.~3, pp.~211--252, 2015.

\bibitem{rabiner1989tutorial}
L.~R. Rabiner, ``A tutorial on hidden {Markov} models and selected applications
  in speech recognition,'' {\em Proceedings of the IEEE}, vol.~77, no.~2,
  pp.~257--286, 1989.

\bibitem{kendall1945}
M.~G. Kendall, ``The treatment of ties in ranking problems,'' {\em Biometrika},
  vol.~33, no.~3, pp.~239--251, 1945.

\end{thebibliography}
\end{document}